\documentclass[]{pasj02} 
\usepackage[switch,mathlines]{lineno} 

\jyear{2024}
\Received{}
\Accepted{}

\begin{document} 

\title{Three-Dimensional Expansion of Iron Ejecta in Cassiopeia A with Chandra and XRISM}

\author{
Aya \textsc{Bamba},\altaffilmark{1,2,3}\orcid{0000-0003-0890-4920}
\email{bamba@phys.s.u-tokyo.ac.jp}
Tomoki \textsc{Kadoya},\altaffilmark{1}
Kouichi \textsc{Hagino},\altaffilmark{1}\orcid{0000-0003-4235-5304}
Jacco \textsc{Vink},\altaffilmark{4,5,1}\orcid{0000-0002-4708-4219}
Manan \textsc{Agarwal},\altaffilmark{4,1}\orcid{0000-0001-6965-8642}
Masahiro \textsc{Ichihashi},\altaffilmark{1}\orcid{0000-0001-7713-5016}
Paul \textsc{Plucinsky}\altaffilmark{6}\orcid{0000-0003-1415-5823}
\altaffiltext{1}{Department of Physics, Graduate School of Science, The University of Tokyo, 7-3-1 Hongo, Bunkyo-ku, Tokyo 113-0033, Japan}
\altaffiltext{2}{Research Center for the Early Universe, School of Science, The University of Tokyo, 7-3-1 Hongo, Bunkyo-ku, Tokyo 113-0033, Japan}
\altaffiltext{3}{Trans-Scale Quantum Science Institute, The University of Tokyo, Tokyo  113-0033, Japan}
\altaffiltext{4}{Anton Pannekoek Institute/GRAPPA, University of Amsterdam, Science Park 904, 1098 XH Amsterdam, The Netherlands}
\altaffiltext{5}{SRON Netherlands Institute for Space Research, Niels Bohrweg 4, 2333 CA Leiden, The Netherlands}
\altaffiltext{6}{Harvard-Smithsonian Center for Astrophysics, MS-3, 60 Garden Street, Cambridge, MA, 02138, USA}
}




\maketitle

\begin{abstract}
Measurement of the expansion structure of ejecta in supernova remnants (SNRs) provides important information about their explosion mechanism. 
We combine Chandra proper-motion measurements with XRISM Doppler measurements to derive an approximate, region-averaged three-dimensional velocity map of the Fe-K-emitting ejecta in Cassiopeia A (Cas A), one of the youngest Galactic core-collapse SNRs. 
Proper motions were measured using the optical flow method. 
We find that the Fe-K-emitting ejecta expand predominantly outward, except in the western region, with projected velocities of up to $\sim4580$ km s$^{-1}$. 
The southeastern ejecta exhibit the highest three-dimensional velocity, reaching $5160\pm320$ km s$^{-1}$, based on the combined proper-motion and Doppler measurements. 
This velocity exceeds those measured for intermediate-mass elements (IMEs) such as Si and S, and the corresponding ejecta exhibit a relatively small line-of-sight velocity dispersion. 
These results are broadly consistent with the previously suggested inversion between the Fe-rich and IME-rich ejecta layers. 
The apparent inward motion or emergence of a new Fe-K emission component in the western region may reflect interaction with dense surrounding material, 
although the present data do not allow us to distinguish uniquely between these possibilities or to determine their physical origin.
\end{abstract}


\section{Introduction}

Supernovae and their remnants (supernova remnants; SNRs) supply heavy elements into interstellar space, which leads to the chemical evolution of the universe.
Measurements of the expansion of the ejecta of SNRs show how these heavy elements are ejected.
It also provides insight into the explosion mechanism of the supernova itself,
because asymmetric effects are essential for the explosion mechanism of core-collapse supernovae \citep[for example]{wongwathanarat2015,orlando2022,burrows2024}.

Cassiopeia A (Cas A) is one of the youngest Galactic core-collapse SNRs.
The remnant has a radius of $\sim$2.8~arcmin,
corresponding to about 2.8~pc at a distance of 3.4~kpc \citep{reed1995}.
Given its young age and small distance, Cas A is an ideal target for studying its expansion and explosion mechanisms.
The shock expansion has been measured using proper motion studies of the synchrotron X-ray emission \citep{vink2022}.
\citet{sato2018} conducted an expansion study with the synchrotron-dominated X-ray band using the optical flow method \citep{farneback2003}.
They also measured the expansion of Si ejecta using the Si K-line band emission, since thermal X-ray emission from Cas A is dominated by emission from ejecta.
\citet{tsuchioka2022} measured the proper motion of the Fe-ejecta-dominant region in the southeastern region of Cas A.
\citet{sakai2024} also performed a proper-motion study using a multiepoch maximum likelihood estimation approach.
Note that these studies used a wide-band energy range, not Fe-K band.
For the line-of-sight expansion, \citet{vink2025} generated a Doppler shift map of Si and S emission lines with the excellent energy resolution of Resolve onboard X-Ray Imaging and Spectroscopy Mission (XRISM).
\citet{suzuki2025} also performed a similar analysis, and measured the three-dimensional velocity of Si ejecta in parts of Cas A, utilizing both Resolve spectroscopy and Chandra imaging analysis.
Given that Fe ejecta are among the most important components for understanding the explosion mechanism, \citet{bamba2025} generated a line-of-sight velocity map for Fe ejecta with Resolve. 
Unlike CCD observations, XRISM/Resolve provides sufficient spectral resolution to resolve the Fe-K line centroid with much smaller uncertainty from line blending and instrumental broadening.
This enables direct measurements of bulk Doppler velocities of Fe-rich ejecta that were difficult to isolate with previous CCD-based observations.
\citet{agarwal2026} compared the Doppler velocities and broadening of intermediate-mass elements and Fe-group elements with XRISM/Resolve at $1' \times 1'$ resolution. They report that the differences in Doppler properties are largest near the center and decrease radially outward.
 
Although the Chandra archive has accumulated deep exposures of Cas A, a systematic remnant-wide proper-motion analysis of the Fe-K-emitting ejecta on spatial scales directly comparable to the XRISM/Resolve Doppler measurements has not previously been available. In this work, we measure region-averaged proper motions on spatial scales matched to the XRISM/Resolve observations, providing an approximate remnant-wide three-dimensional velocity map.
Unlike intermediate-mass elements such as Si and S, Fe-rich ejecta originate from the deepest layers of the progenitor and therefore provide a more direct probe of the innermost explosion dynamics and large-scale overturning during the supernova explosion.
A remnant-wide kinematic study of Fe-K-emitting ejecta is therefore essential for understanding the asymmetric transport and mixing of the innermost ejecta layers.

The three-dimensional reconstruction of Cas~A has a long history in optical studies of fast-moving knots \citep[for example]{reed1995,fesen2006, delaney2010}.
Although the spatial distribution and Doppler properties of the Fe-rich ejecta have been investigated extensively, their remnant-wide three-dimensional kinematic relation to the optical ejecta has not yet been systematically characterized.
Here we present an approximate remnant-wide three-dimensional velocity map of the Fe-K-emitting ejecta by combining Chandra proper motions with XRISM Doppler measurements, 
providing a framework for future comparison.

In this paper, we present the proper motion map of the X-ray-emitting Fe ejecta with Chandra using Fe-K band only for the first time, and construct an approximate region-averaged three-dimensional expansion map together with the Resolve results by \citet{bamba2025}.
Chandra imaging alone can measure projected expansion velocities, whereas line-of-sight velocities of the Fe-rich ejecta have previously been inferred from X-ray spectroscopy. 
XRISM substantially improves the robustness of Fe-K Doppler velocity measurements through its high spectral resolution.
The combination of Chandra proper motions and XRISM Doppler measurements provides a common observational framework for investigating the approximate large-scale three-dimensional kinematics of the Fe-rich ejecta and facilitates future systematic comparisons with the optical knot population,
although the angular scale of the ejecta structure is much smaller than the Resolve point spread function, and as a result each XRISM region likely contains multiple velocity components with different Doppler shifts.
In Section~\ref{sec:obs}, we summarize the Chandra and XRISM observations of Cas A.
Section~\ref{sec:chandra} and Section~\ref{sec:3D} show the results on the proper motion and three-dimensional measurement with the Fe K complex.
Section~\ref{sec:discuss} is devoted to the discussion on the comparison with other measurements and explosion models.
Throughout this paper, we quote errors at the 1$\sigma$ confidence level.

\section{Observations and Data Reduction}
\label{sec:obs}

\subsection{Chandra}

To study the evolution of Fe knot morphology,
we selected data obtained with the ACIS-S detector \citep{garmire2003,bautz1998} onboard Chandra \citep{weisskopf1996,weisskopf2000,weisskopf2003} with exposure times longer than 10~ks.
As shown in Table~\ref{tab:obslog}, 26 observations were selected in a 19-year interval.
We reprocessed the data using \texttt{CIAO} 4.17.0 \citep{fruscione2006} and the calibration database (\texttt{CALDB}) version 4.12.0.
For the spectral analysis, we used {\tt XSPEC} version 12.14.1 \citep{arnaud1996} in HEASoft 6.34.


\begin{table}
  \tbl{Chandra observation log.}{%
  \begin{tabular}{ccc}
      \hline
  ObsID & Obs start & Exposure \\ 
 & (yyyy/mm/dd) & (ks) \\ 
  \\ 
      \hline
00114 & 2000/01/30 & 49.93 \\
01952 & 2002/02/06 & 49.66 \\
04634 & 2004/04/28 & 148.63 \\
04635 & 2004/05/01 & 135.04 \\
04636 & 2004/04/20 & 143.48 \\
04637 & 2004/04/22 & 163.49 \\
04638 & 2004/04/14 & 164.53 \\
04639 & 2004/04/25 & 79.05 \\
05196 & 2004/02/08 & 49.53 \\
05319 & 2004/04/18 & 42.26 \\
05320 & 2004/05/05 & 54.37 \\
09117 & 2007/12/05 & 24.84 \\
09773 & 2007/12/08 & 24.84 \\
10935 & 2009/11/02 & 23.26 \\
10936 & 2010/10/31 & 32.24 \\
12020 & 2009/11/03 & 22.38 \\
13177 & 2010/11/02 & 17.24 \\
14229 & 2012/05/15 & 49.09 \\
14480 & 2013/05/21 & 48.77 \\
14481 & 2014/05/12 & 49.42 \\
14482 & 2015/04/30 & 49.42 \\
18344 & 2016/10/21 & 25.75 \\
19604 & 2017/05/16 & 49.53 \\
19605 & 2018/05/15 & 49.42 \\
19606 & 2019/05/13 & 49.42 \\
19903 & 2016/10/20 & 24.65 \\
      \hline
    \end{tabular}}\label{tab:obslog}
\begin{tabnote}
\end{tabnote}
\end{table}

\subsection{XRISM}

For the Doppler measurement of the Fe-K emission,
we used results obtained with Resolve \citep{ishisaki2022,ishisaki2025,kelley2025} onboard XRISM \citep{tashiro2025}.
Cas A has been observed with XRISM
in two pointings, southeast pointing (ObsID: 000129000, hereafter SE) and northwest pointing (ObsID: 000130000, hereafter NW),
with exposures of 181.3~ks for the SE and 166.6~ks for the NW.
\citet{bamba2025} generated a Doppler map of the Fe-K complex,
which we use to construct an approximate region-averaged 3D velocity map of the Fe-K complex in Cas A.

\section{Proper motion measurement with Chandra}
\label{sec:chandra}

In this section, we show how we derive the proper motion of Fe ejecta.
As the first step, the Fe-K complex images in each observation were created, as shown in section~\ref{sec:energyband}.
The optical flow method was applied to measure the proper motion and the results were compiled into the proper-motion map
(see Section~\ref{sec:opticalflow}).

\subsection{Making Fe-K complex images}
\label{sec:energyband}

\begin{figure}
 \begin{center}
  \includegraphics[width=7cm]{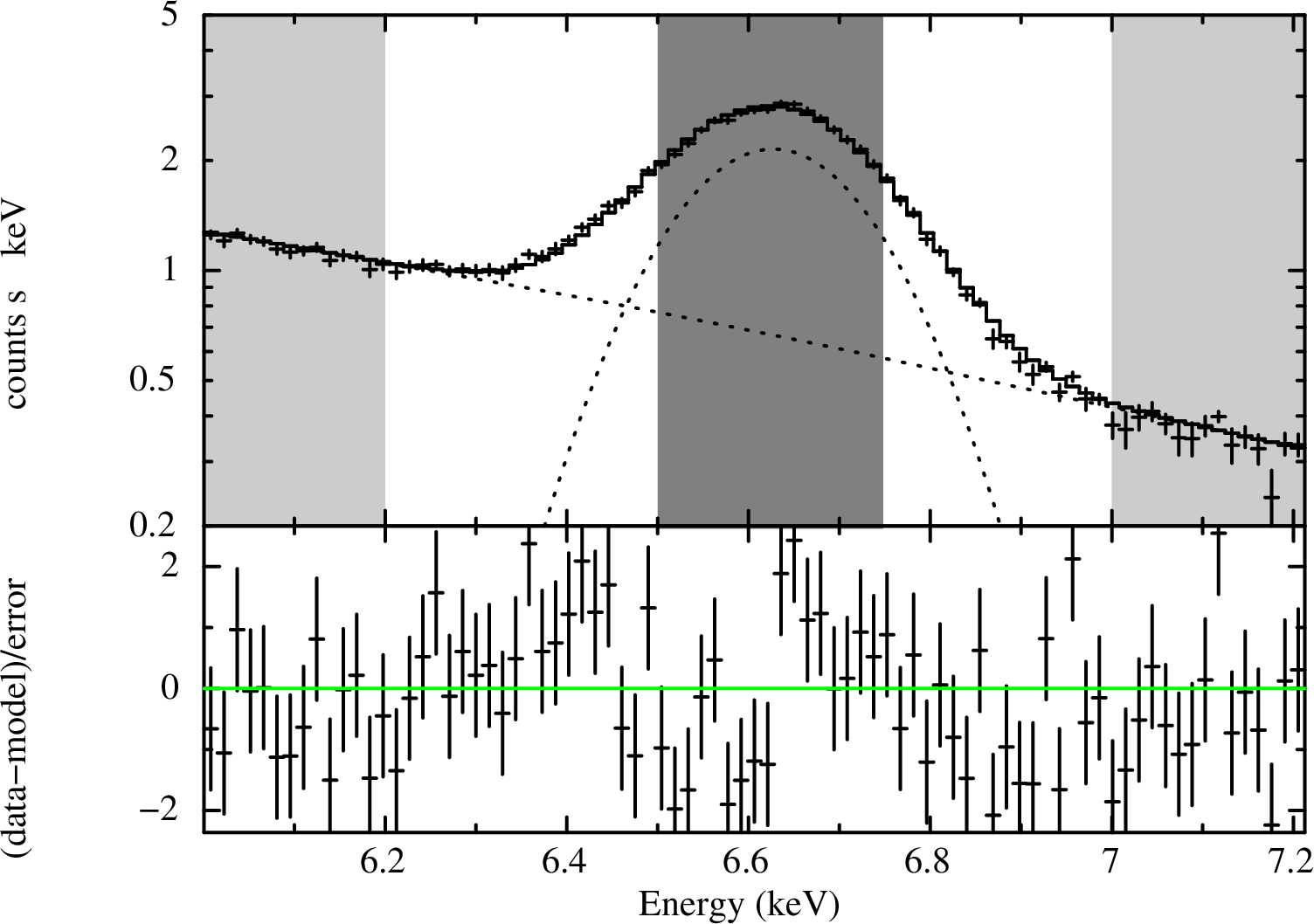} 
\end{center}
\caption{Upper panel: The Fe-K band spectrum of the entire remnant. Dotted lines represent power-law and Gaussian components. Dark and light gray regions show the definition of Fe-K complex-dominant energy band and continuum-dominant energy band, respectively. Lower panel: Residuals from the best-fit model.
{Alt text: A line graph showing Fe-K band spectrum and the energy bands we used for the imaging analysis.}
}\label{fig:spec}
\end{figure}

Figure~\ref{fig:spec} shows 6.0--7.2~keV-band spectrum from the entire remnant in ObsID=4638, the longest observation.
Background photons were accumulated from a source-free region.
A strong Fe-K complex is visible, which was used by \citet{bamba2025} for a Doppler velocity study with XRISM.
From this spectrum, we selected a line-dominant energy band as 6.5--6.75~keV, and continuum-dominant energy bands as 6.0--6.2~keV and 7.0--7.2~keV.
The images in these energy bands were extracted from each dataset.

\begin{figure}
 \begin{center}
 \includegraphics[width=7cm]{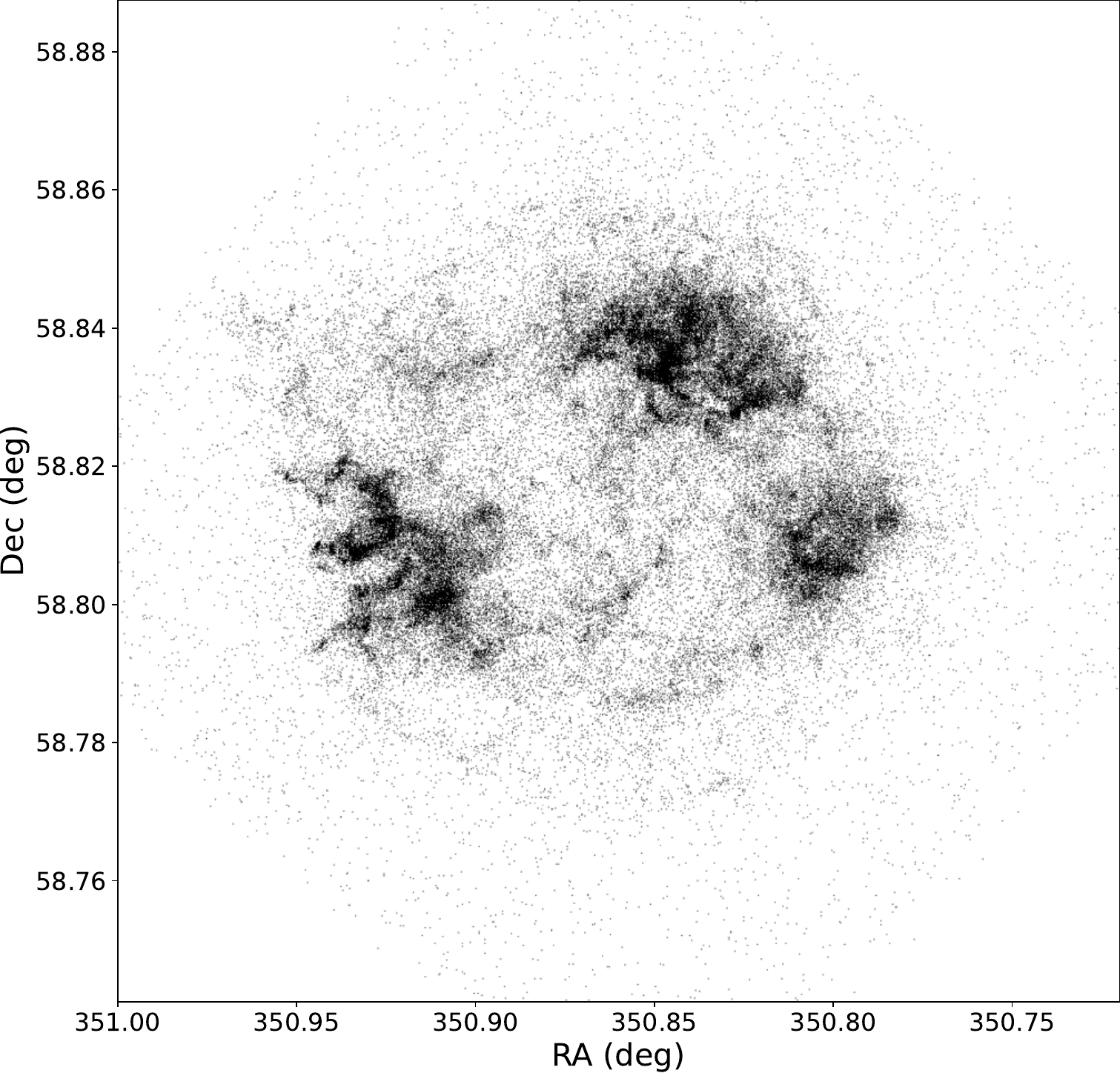} 
\end{center}
\caption{The background-subtracted image in 6.5--6.75~keV band for ObsID=4638, showing Fe-K complex image.
Color scale is in linear, and the binned with 0.5 arcsec.
{Alt text: An image of Fe-K complex.}
}\label{fig:FeK-image}
\end{figure}

To extract Fe-K complex images, an estimate of the continuum contribution to the line-dominant band is required.
We fitted the spectra with a power-law plus a Gaussian model, and faked the spectrum with the best-fit parameters to estimate the count rate in the Fe-K complex-dominant energy band.
The count rates of the continuum component in the continuum-dominant energy band and Fe-K complex dominant energy band are 0.252 counts~s$^{-1}$ (in 6.0--6.2~keV), 0.078 counts~s$^{-1}$ (in 7.0--7.2~keV), and 0.175 counts~s$^{-1}$ (in 6.5--6.75~keV, faked value), respectively.
We thus subtracted continuum-dominant-band images with the factor of $0.175/(0.252+0.078) = 0.53$, from the Fe-K complex-dominant images to derive the Fe-K complex images.
For pixels with negative values due to over-subtraction, we set their values to 0.
We set the intensity scale from 0 to 255 in all images.
Figure~\ref{fig:FeK-image} shows the derived Fe-K complex image for ObsID=4638. 
One can see that there is no clear thin filament surrounding the remnant, which predominantly arises from synchrotron continuum emission from high-energy electrons accelerated on the shock fronts \citep{vink2003,bamba2005}.

\subsection{Measuring proper motion}
\label{sec:opticalflow}

\begin{figure}
    \begin{center}
        \includegraphics[width=8cm]{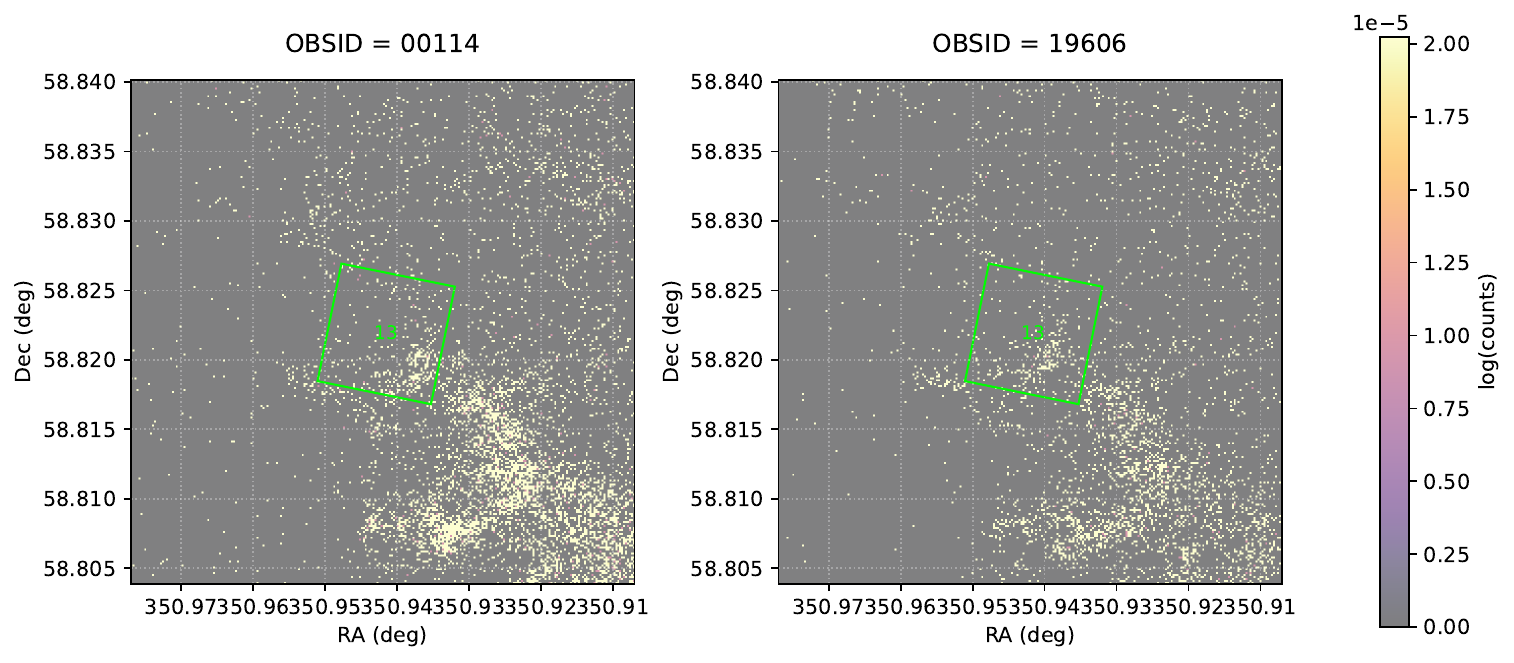}
\end{center}
\caption{The Fe-K complex images around pix 2 in the XRISM SE observation, taken on 2000 Jan. 30 (ObsID=114, left panel) and 2019 May 13 (ObsID=19606, right panel).
These images are binned with 0.5 arcsec.
{Alt text: Fe-K complex images at different epochs.}
}\label{fig:example}
\end{figure}

Figure~\ref{fig:example} shows the comparison of Fe-K emission around pix 2, in the XRISM SE observation taken on 2000 Jan. 30 (ObsID=114, left panel) and 2019 May 13 (ObsID=19606, right panel).
One can see that the Fe knot is moving roughly from west to east.

Here, we applied the optical flow method to quantitatively measure the motion.
For each observation, the displacement field was measured relative to ObsID=4638 (obtained in 2004) using the optical-flow method. The resulting positional offsets were then fitted as a function of time with a linear model to derive the proper motions.
The effective temporal baseline is approximately 16 years for the latest observations.
This multi-epoch fitting reduces the impact of fluctuations associated with individual epoch pairs.
The optical flow algorithm by \citet{farneback2003} is applied,
which is the same as that in \citet{sato2018}.
We adopted the parameter set of \citet{sato2018} because it provides spatial smoothing scales comparable to the characteristic angular size of the Fe-K ejecta structures while maintaining sufficient statistical stability in the low-count Fe-K images.
The time difference $(dt)$ between data sets is taken to be the starting point of the observations listed in Table~\ref{tab:obslog}.  The displacement in the east-west direction $(dx)$ and in the north-south direction $(dy)$ is determined by the optical flow method for each position in every observation.
The image from ObsID=4638 with the bin size of $0.5
\times 0.5$~arcsec is used for the reference of the displacement.
We also performed the same analysis with the bin size of $1\times 1$~arcsec and found basically no difference.
We calculated $dx$ and $dy$ for every 0.5 arcsec $\times$ 0.5 arcsec regions.
After obtaining the results, we averaged them over the XRISM/Resolve pixel size.

Since the optical-flow algorithm itself does not provide formal statistical uncertainties for individual vectors, 
we characterized the regional scatter of the proper-motion vectors within each XRISM region using the dispersion of the measured proper-motion vectors within that region. 
This quantity is not interpreted as a formal fitting uncertainty derived from the optical-flow algorithm. 
Rather, it represents an empirical estimate of the regional scatter of the measured proper motions, ranging from 0.002 to 0.037 arcsec. 
The regional scatter may include contributions from measurement uncertainty, intrinsic sub-regional velocity variations, morphological complexity, and the regularization of the optical-flow algorithm.
In addition, we should also include the astrometric uncertainty.
CIAO provides the astrometry correction tool, {\tt fine\_astro}, which can improve the relative astrometry between Chandra observations by comparing the positions of point sources common to both observations.
Unfortunately, point sources cannot be reliably identified in the Cas A field due to the extremely bright and variable diffuse emission from Cas A itself. 
We therefore rely upon the aspect reconstruction provided by the standard CIAO processing which provides an absolute astrometric accuracy of 0.66 arcsec ($1.0\sigma$ radius). for observations on the S3 detector within 3.0 arcmin of the aimpoint\footnote{(https:\/\/cxc.harvard.edu\/cal\/ASPECT\/celmon\/\#offset\_history)}. 
We combined the estimated regional scatter and the astrometric uncertainty in quadrature and adopted the resulting value as the uncertainty of the region-averaged proper motion.
Note that the estimated regional scatter is generally much smaller than the astrometric uncertainty.

The absolute astrometric uncertainty of 0.66~arcsec corresponds to a systematic velocity uncertainty of approximately 660~km~s$^{-1}$ over a 16-year baseline.
Thus, velocities below 660~km~s$^{-1}$ should be interpreted with caution, as they are comparable to the systematic astrometric uncertainty.

\begin{figure}
    \begin{center}
        \includegraphics[width=7cm]{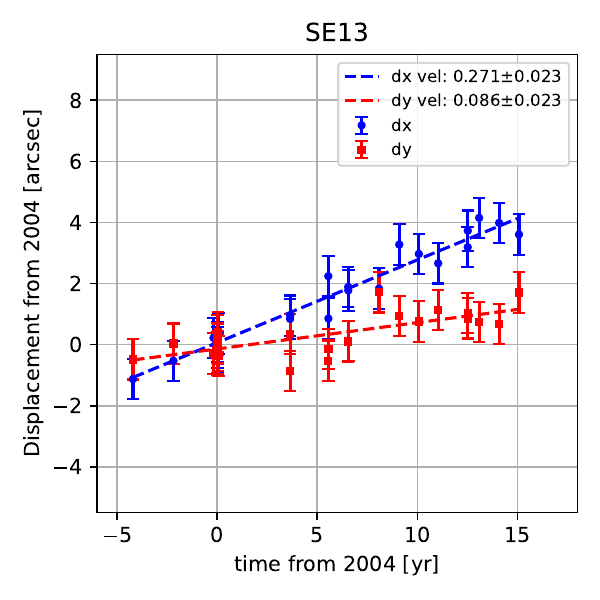}
    \end{center}
\caption{An example of $dt$ vs. $dx$ (blue) and $dy$ (red) in the XRISM/Resolve pix 13 in the SE observation.
Dashed lines represent the best-fit line of the displacement.
The error bars represent the adopted uncertainties of the region-averaged proper motions, obtained by combining the estimated regional vector scatter with the astrometric uncertainty.
{Alt text: A line graph.}}\label{fig:dxdy_example}
\end{figure}

Figure~\ref{fig:dxdy_example} shows an example of $dt$ vs. $dx$ and $dy$ for the pixel 13 in the XRISM SE observation \citep{plucinsky2025}.
Errors include both the regional-scatter and astrometric uncertainties.
In this case, $dx$ increases with time, whereas $dy$ keeps almost constant,
implying that Fe-K emission in this region moves roughly from west to east.
We fitted $dx$ and $dy$ with a linear function of time which is also shown in Figure~\ref{fig:dxdy_example}, and derived the best-fit $dx/dt$ and $dy/dt$.
In this case, $dx/dt = 4370\pm380$~km~s$^{-1}$ and $dy/dt = 1390\pm380$~km~s$^{-1}$, thus the two-dimensional velocity is determined to be $4580\pm380$~km~s$^{-1}$.
Pix 13 actually shows one of the largest velocities among all pixels.
The results for the other XRISM/Resolve pixels are summarized in Appendix A.

\begin{figure}
 \begin{center}
  \includegraphics[width=8cm]{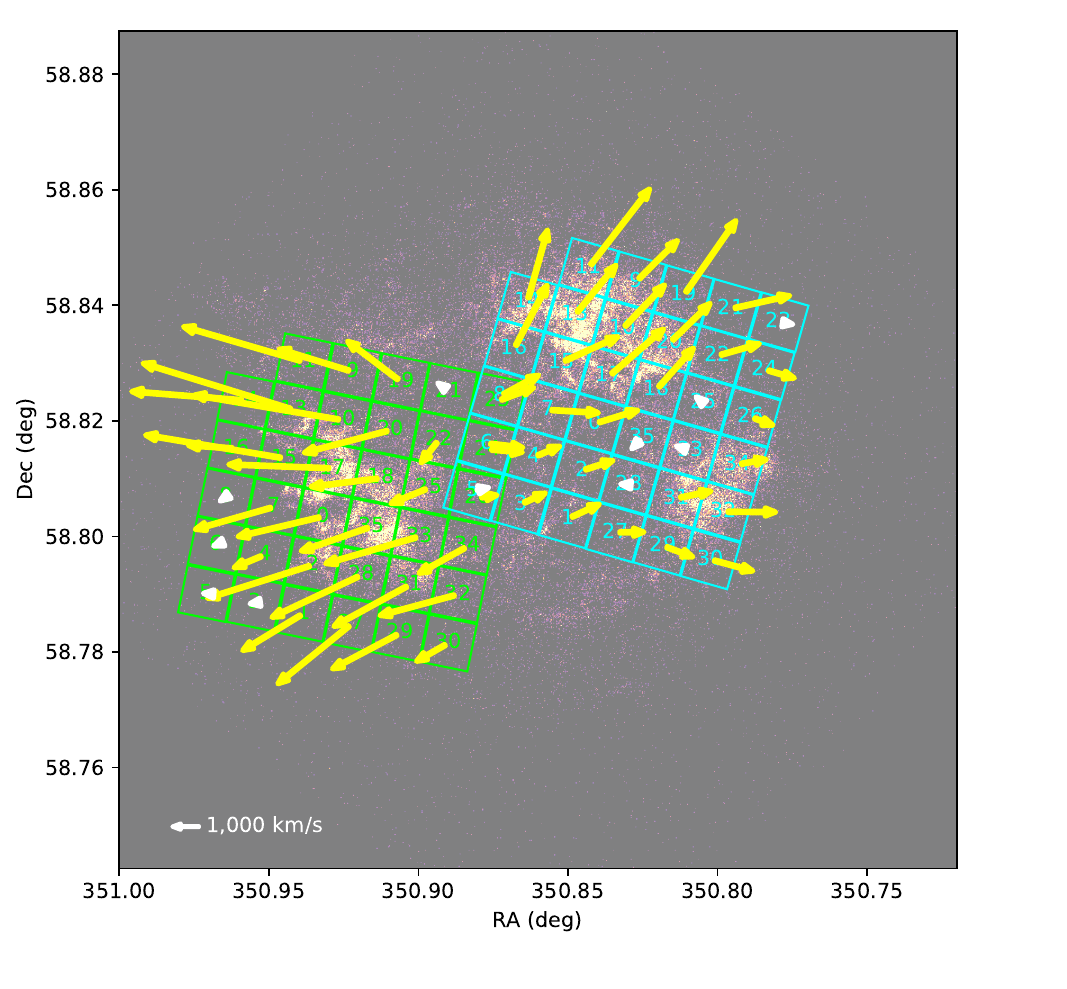} 
\end{center}
\caption{The proper motion map of Fe-K complex (arrows) overlaid on the Fe-K complex intensity map.
Arrow lengths represent the magnitude of the projected expansion velocity.
Yellow arrows indicate velocities larger than the estimated systematic astrometric uncertainty (660~km~s$^{-1}$), while white arrows indicate velocities comparable to or smaller than this uncertainty.
The images are binned with a pixel size of 0.5~arcsec.
{Alt text: An image showing proper motion of Fe-K complex.}}\label{fig:FeK-proper}
\end{figure}

The derived $dx/dt$ and $dy/dt$ are the proper motion velocities in east-west direction and north-south direction in each XRISM/Resolve pixel, respectively.
We thus made the map of proper motion, which can be seen in Figure~\ref{fig:FeK-proper}.
Yellow arrows indicate velocities larger than the estimated systematic astrometric uncertainty (660~km~s$^{-1}$), while white arrows indicate velocities comparable to or smaller than this uncertainty.
The derived motions show coherent large-scale expansion patterns rather than random pixel-scale fluctuations, supporting the robustness of the adopted parameter set.

We made the same analysis with 1~arcsec binning and 2~arcsec binning images, which returned basically the same results.
A systematic error in the subtraction of the continuum component may affect our results. 
We therefore also performed the same analysis with 10\% larger or smaller continuum level, and no significant difference is found in our results with the difference of only less than 0.05~arcsec/16~years, which is negligible compared with other error factors.

In order to assess the robustness of the proper-motion measurements against astrometric uncertainties, we additionally performed an independent multi-epoch radial-expansion analysis (see Appendix B), which iteratively solves for inter-epoch positional offsets. The resulting velocities are broadly consistent with those obtained from the optical-flow method.
This analysis serves as an independent consistency check of the optical-flow results.

\subsection{Three-dimensional kinetic motion of Fe ejecta}
\label{sec:3D}

In the last step, we combined the proper motion map derived in the previous sections with the Doppler shift map of Fe-K complex derived with XRISM/Resolve \citep{bamba2025}.
The three-dimensional velocity of the Fe line emission in SE pix 11 is now $5160\pm320$~km~s$^{-1}$, which is the largest among all the pixels.

\begin{figure}
 \begin{center}
  \includegraphics[width=9cm]{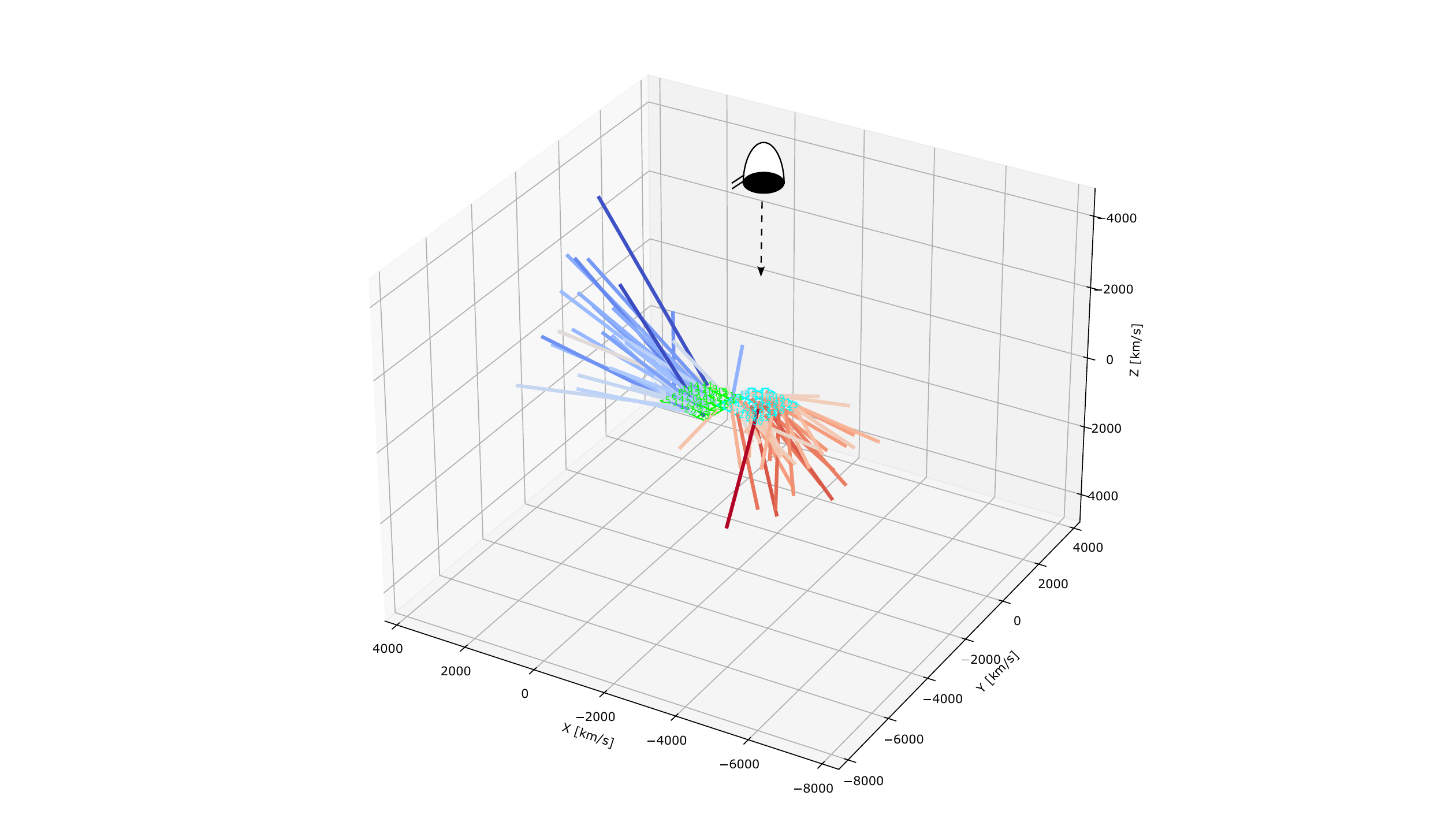} 
\end{center}
\caption{The approximate region-averaged three-dimensional motion map of the Fe-K complex (arrows) with Resolve field of view.
The arrow length represents the velocity magnitude, while the color indicates the line-of-sight Doppler velocity adopted from XRISM/Resolve measurements.
X, Y, and Z represent east-west, north-south, and line-of-sight directions, respectively.
The displayed velocities should be interpreted as approximate region-averaged quantities because of the broad Resolve PSF.
Cas~A is viewed from above, as indicated by the eye symbol.
{Alt text: An image showing three-dimensional motion of Fe-K complex.}
}\label{fig:FeK-3D}
\end{figure}

Figure~\ref{fig:FeK-3D} provides the first remnant-wide view of the large-scale kinematics of Fe-K-emitting ejecta in Cas A by combining projected expansion velocities with XRISM Doppler measurements.
The resulting velocity structure suggests that the Fe-rich ejecta preserve large-scale asymmetries originating from the innermost supernova explosion.
Note that the Fe complex is broadened by a mixture of blueshifted and redshifted components.
Our three-dimensional velocity estimate ignores this effect; and thus does not provide a completely accurate representation.
However, iron ejecta in the southeastern region have a narrow and blueshifted feature \citep{bamba2025},
implying that the ejecta actually move toward us
with the velocity derived using our method.

\section{Discussion}
\label{sec:discuss}

We successfully performed a quantitative measurement of the three-dimensional motion of Fe ejecta in Cas A
using both Chandra and XRISM data.
As shown in Figure~\ref{fig:FeK-proper} and Figure~\ref{fig:FeK-3D},
the iron ejecta expand outward in most directions. 
The three-dimensional expansion velocity is up to $\sim$5160~km~s$^{-1}$, depending on the direction.

In this section, we compare our results with those obtained at other wavelengths.
Chandra imaging measures only projected proper motions. Line-of-sight velocities have previously been inferred from X-ray spectroscopy, but XRISM substantially improves the robustness of Fe-K Doppler measurements through its unprecedented spectral resolution.
The combination of Chandra proper motions and XRISM Doppler measurements 
provides a common observational framework for future systematic comparisons of the large-scale three-dimensional kinematics of the X-ray Fe-rich ejecta with those of the optical knot population across the remnant.
However, the characteristic angular scale of the ejecta structures in Cas A is significantly smaller than the Resolve PSF, and each XRISM extraction region likely contains multiple kinematic components with different Doppler velocities. Therefore, the Doppler velocities adopted in this work should be interpreted as region-averaged quantities rather than velocities of individual ejecta clumps. In addition, the Chandra proper motions and XRISM Doppler velocities are not necessarily associated with identical substructures because of the different spatial resolutions and PSF mixing. Thus, the derived three-dimensional velocity structure should be regarded as an approximate large-scale kinematic picture of the Fe-rich ejecta.

Previous X-ray studies have shown that Fe-rich ejecta in Cas~A are distributed in distinct clumps and often located outside of the intermediate-mass elements, suggesting substantial mixing of inner ejecta layers during the explosion \citep[for example]{hwang2012,tsuchioka2022}.
In addition, optical studies have reconstructed the three-dimensional structure of high-velocity knots and jet-like features, and reveal strong asymmetries in the outer ejecta \citep[for example]{milisavljevic2013}.
Our results confirm these multi-wavelength studies not only by X-ray proper motion study only but also by providing approximate region-averaged three-dimensional velocity map.

\begin{figure}
    \begin{center}
        \includegraphics[width=8cm]{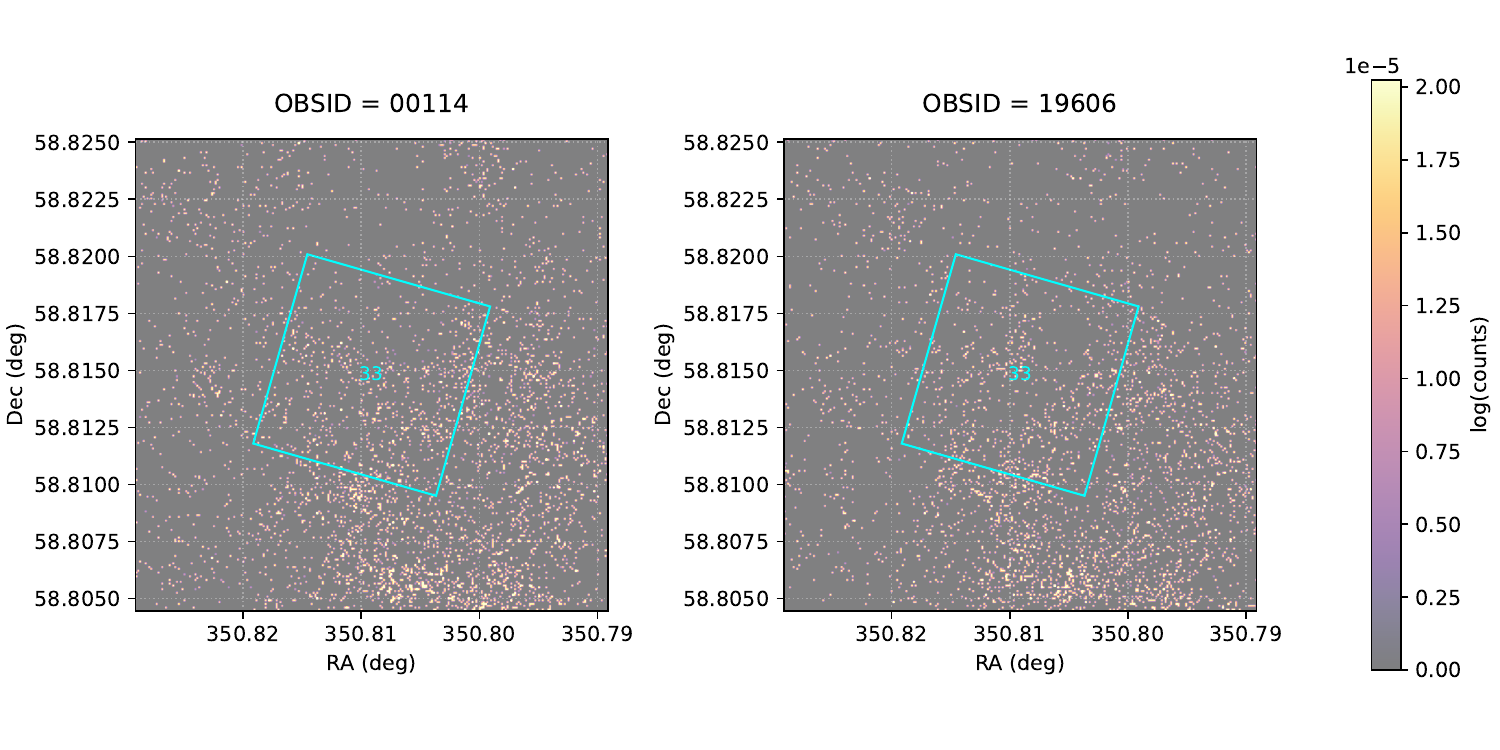}
    \end{center}
    \caption{Fe-line image around the NW pix 33 region taken from ObsID=114 (left) and ObsID=19606 (right). Blue boxes show the Resolve pix 33 region in the both panels. 
    {Alt text: Fe-line images in different epochs.}}\label{fig:pix33}
\end{figure}


\citet{sato2018} measured the time evolution of the Si and continuum band images with the optical flow method.
Both show overall expansion, whereas the western and southern rims show significant inward motion in the continuum band.
This feature is also noticed by \citet{vink2022}.
This inward shock may result from interaction with dense material \citep{kilpatrick2014}, although \citet{zhou2018} did not find a clue of the interaction.
From Figure~\ref{fig:FeK-proper}, we also find that the iron ejecta also show the possible inward movement in the western and southern regions.
We have checked whether the ejecta really show inward motion or not with the raw images.
Figure~\ref{fig:pix33} shows the comparison of two epochs around NW pix33 region.
We can see that there is some difference, but it is not clear whether it is a parallel shift of steady emission component or the emergence of a new emission component.
Both cases imply the interaction with dense material
with the inward movement or the larger emission measure.
Interestingly, this region also coincides with the dense material  called the "green monster" discovered by James Webb Space Telescope (JWST)
\citep{milisavljevic2024,vink2024}.
The interpretation of the western region, including the possible inward motion or the emergence of a new component, remains somewhat tentative. Given the relatively low count statistics in this region, as well as potential uncertainties associated with continuum subtraction, morphological evolution, and the spatial mixing inherent to the XRISM/Resolve PSF, the current data do not allow us to uniquely distinguish between these scenarios. Therefore, the connection to interaction with dense material, including a possible association with the JWST green monster, should be regarded as a plausible but not yet firmly established interpretation.
A more detailed comparison with iron ejecta movement and green monster will be done by NewAthena \citep{cruise2025}.

Fast proper motion or line-of-sight movement of iron ejecta is observed especially in the southeastern region
both in this work and previous studies \citep[for example]{tsuchioka2022,bamba2025}.
In this work, we showed for the first time that the pix 11 in the XRISM SE observation shows the fastest three-dimensional velocity,
$5160\pm320$~km~s$^{-1}$.
\citet{suzuki2025} derived three-dimensional velocities in Si band for several $2\times 2$ pixels ("superpixels").
"SE2" superpixel, which contains pixels SE9, 10, 19, and 20, shows $4000\pm 400$~km~s$^{-1}$.
On the other hand, the iron three-dimensional velocity in SE10, the brightest one in the Fe band, is $4960_{-200}^{+360}$~km~s$^{-1}$,
which is significantly larger than Si velocity.
This provides further confirmation of the inversion of iron and Si ejecta.
It is also consistent with the fact that \citet{agarwal2026} found that iron group elements show broader emission lines compared with intermediate-mass elements in most regions.
Note that the velocity estimation in \citet{suzuki2025} is for larger area, so the velocity can be smaller due to the mixing of different velocity components.
Fe ejecta in these regions also show small Doppler broadening (less than 1000~km~s$^{-1}$; \cite{bamba2025}), 
which also implies that Fe ejecta are clumpy.
Although the present comparison is affected by the different spatial samplings of the Fe and IME measurements and by possible PSF mixing in the XRISM data, the overall trend is consistent with the inversion between the Fe-rich and IME ejecta layers suggested in earlier studies.
While previous studies such as \citet{tsuchioka2022} focused primarily on localized southeastern ejecta structures using broad-band X-ray imaging, the present work extends the kinematic investigation to remnant-wide Fe-K-emitting ejecta and incorporates Doppler information from XRISM.

On the other hand, in the northwestern region,
the region with fastest three-dimensional velocity in Si is found in the "NW2" region in \citet{suzuki2025}, where we have NW9, 10, 19 and 20 in our regions,
$5500\pm400$~km~s$^{-1}$,
which is much larger than Fe velocity in our analysis ($\sim 1670-3020$~km~s$^{-1}$)
and the fastest Fe velocity in the entire remnant.
These facts imply that layer inversion occurs only in the southeastern region of Cas A.

The velocity range derived for the Fe-K emitting ejecta (up to $\sim$5160~km~s$^{-1}$) overlaps with those of the optical main-shell knots reported in previous studies \citep[for example]{reed1995}, but is significantly lower than that of the fastest outer optical ejecta ($\sim$14,000~km~s$^{-1}$; \cite{fesen2006}). 
This suggests that X-ray–emitting Fe clumps likely correspond to material in the main shell rather than the most extreme high-velocity optical components. 
A detailed spatial comparison between optical knots and Fe-rich X-ray clumps is beyond the scope of this work. 
However, the present region-averaged velocity map provides a framework for such future comparisons.
Our results are broadly consistent with the large-scale three-dimensional ejecta structure proposed by \citet{delaney2010}.
The present study extends this picture specifically to Fe-K-emitting ejecta by combining projected expansion measurements with high-resolution XRISM Doppler velocities.
In this sense, our analysis provides 
a region-averaged remnant-wide kinematic framework for future systematic comparison between the X-ray Fe-rich ejecta and the optical knot population.
While \citet{delaney2010} discussed the three-dimensional distribution of the X-ray Fe ejecta together with optical and infrared tracers, the present study combines Chandra proper motions with XRISM Fe-K Doppler measurements to provide a framework for future systematic comparison of the large-scale kinematics of the Fe-rich ejecta.

Overall, Figure~\ref{fig:FeK-3D} provides the first remnant-wide view of the large-scale kinematics of Fe-K-emitting ejecta in Cas A. 
The observed velocity structure likely reflects both the intrinsic asymmetry of the supernova explosion and subsequent interaction with the surrounding medium. The eastern ejecta appear to retain signatures of the original explosion geometry, whereas the western ejecta have likely been modified by interaction with the dense ambient material associated with the dense circumstellar material.

\section{Conclusion}
\label{sec:conclusion}

We performed a proper motion study for Fe-K-complex-emitting ejecta.
Using Chandra observations separated by a 16-year interval,
we successfully constructed a proper motion map  of Fe-K line using an optical flow method.
Outward motion was found in almost all directions,
with velocities of up to $\sim$4580~km~s$^{-1}$.
Combined with Doppler velocity measurement obtained with XRISM \citep{bamba2025}, we also constructed  an approximate region-averaged three-dimensional velocity map.
The velocity range is up to $\sim$5160~km~s$^{-1}$,
with the highest velocities in the southeastern region.
In this region, the derived velocity is larger than that of Si,
consistent with the previously suggested Fe/IME layer inversion,
although we need to caution the spectral spatial mixing effect.
On the other hand, the western region shows no significant outward but possible inward motion.
Although it is not clear whether it can be the real inward motion or the emergence of a new emission component, 
both may reflect interaction
with dense material in this region called the "green monster" detected by JWST.
Since the iron line shows a redshift in this region, the dense material may be located behind the remnant.
Future missions such as NewAthena will confirm these results with high sensitivity and better spatial resolution.


\begin{ack}
We thank the anonymous referee for his/her fruitful comments.
We also thank Hirofumi Noda for teaching us the future observatory details.
This research has made use of data obtained from the Chandra Data Archive provided by the Chandra X-ray Center (CXC).
We thank the support by Global Science Graduate Course (GSGC) Project of The University of Tokyo.
This paper employs a list of Chandra datasets, obtained by the Chandra X-ray Observatory, contained in the Chandra Data Collection (CDC) 665 (https:\/\/doi.org\/10.25574/cdc.665).
\end{ack}

\section*{Funding}
This work is supported in part by
the JSPS Core-to-Core Program (grantnumber: JPJSCCA20220002),
and Grants-in-Aid for Scientific Research from the Japanese Ministry of Education, Culture, Sports, Science and Technology (MEXT) of Japan, No.23K25907 (AB).

\appendix 

\onecolumn

\section*{Appendix A: Proper motion measurement results}
\label{sec:propermotion_figures}

\begin{figure}[H]
\begin{center}
\includegraphics[width=2cm]{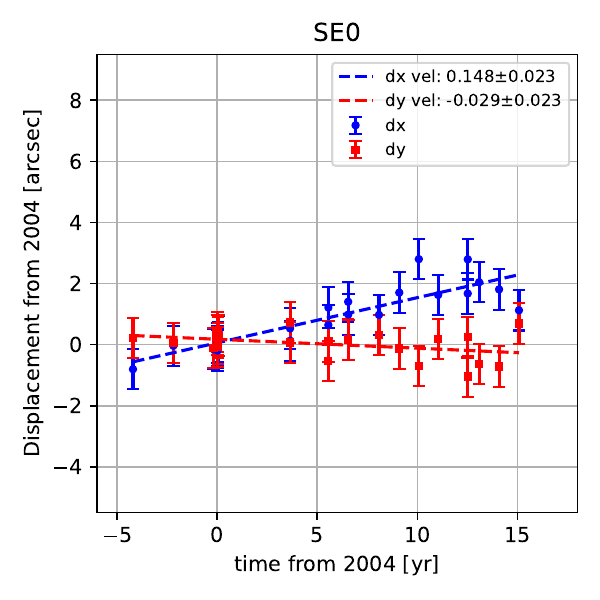}
\includegraphics[width=2cm]{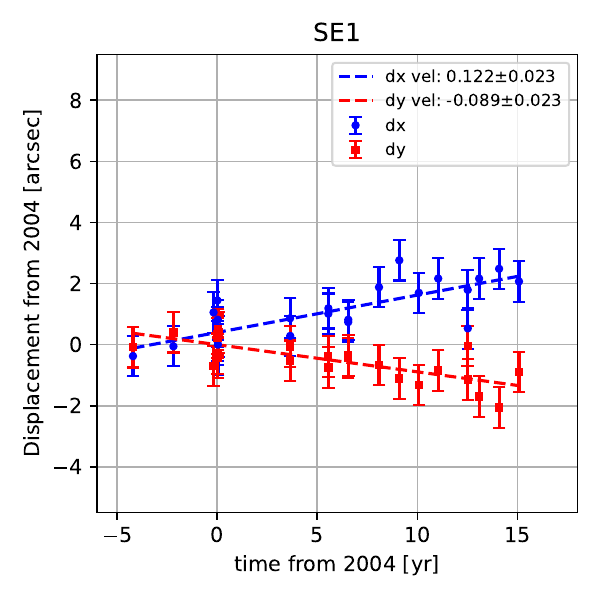}
\includegraphics[width=2cm]{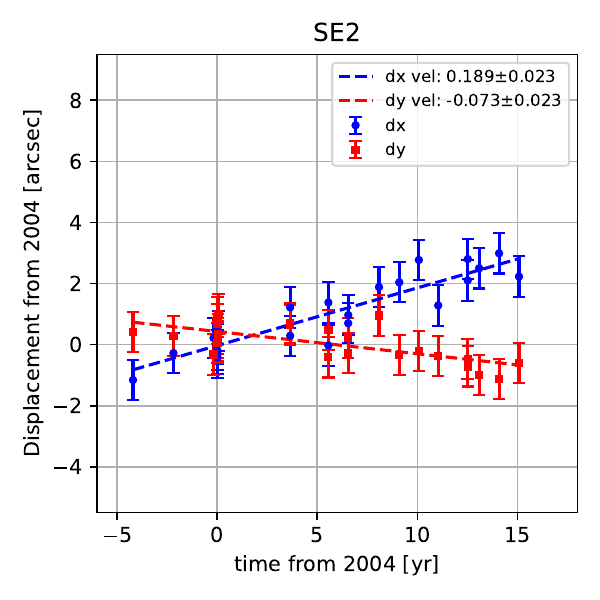}
\includegraphics[width=2cm]{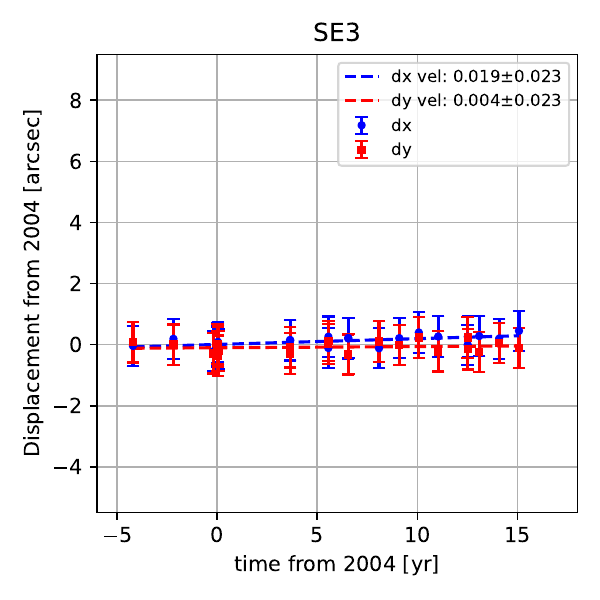}
\includegraphics[width=2cm]{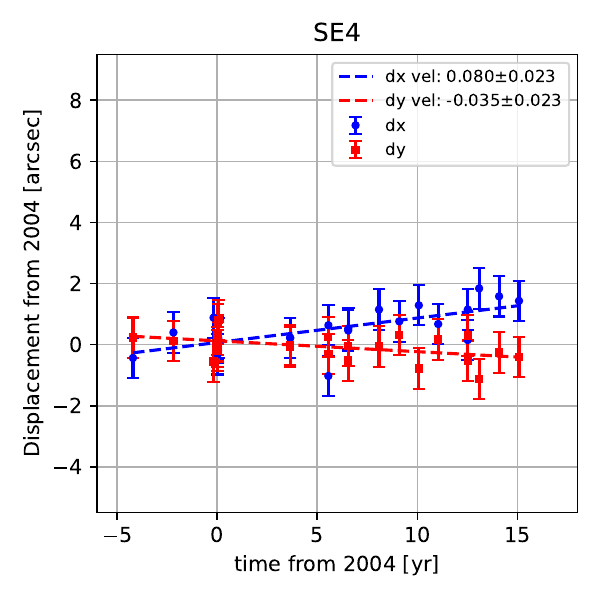}
\includegraphics[width=2cm]{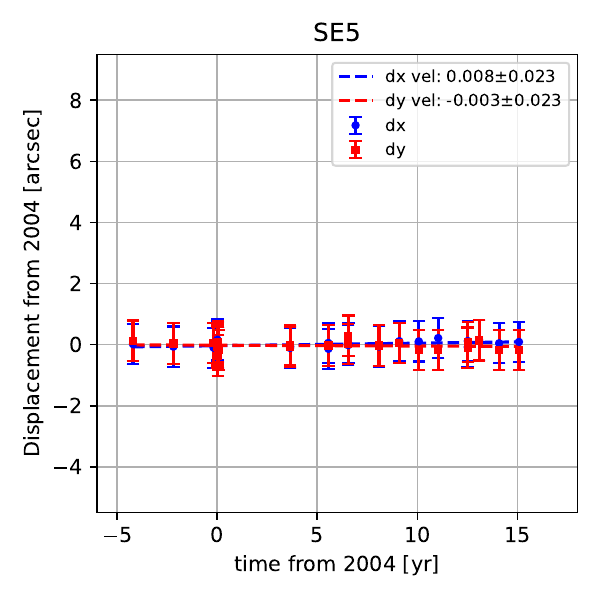}
\includegraphics[width=2cm]{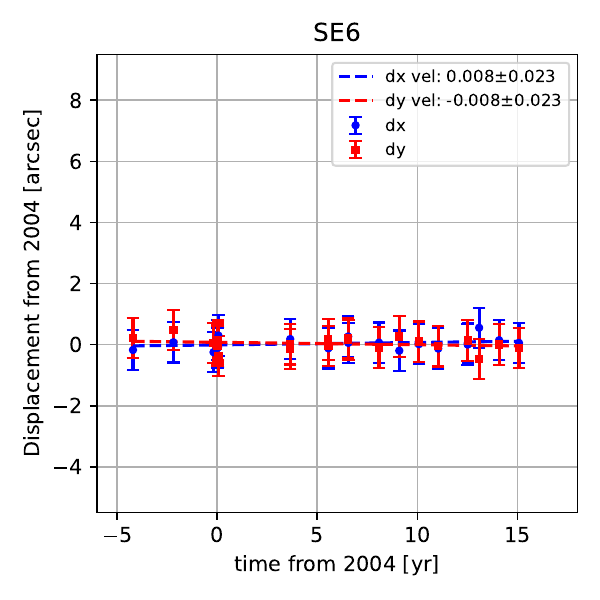}
\includegraphics[width=2cm]{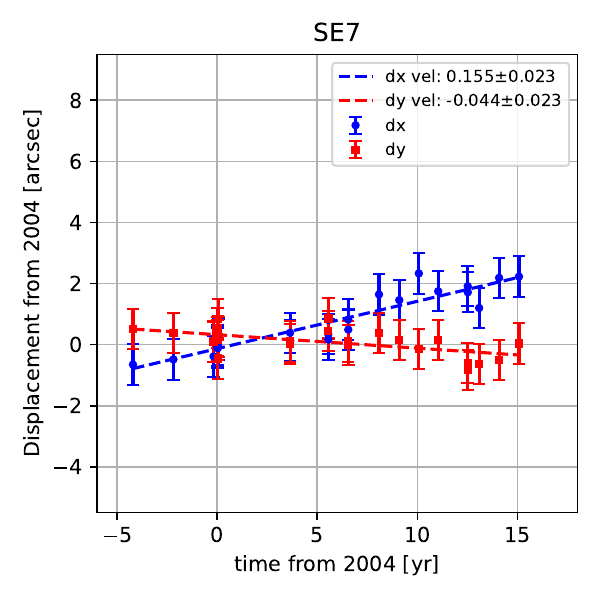}
\includegraphics[width=2cm]{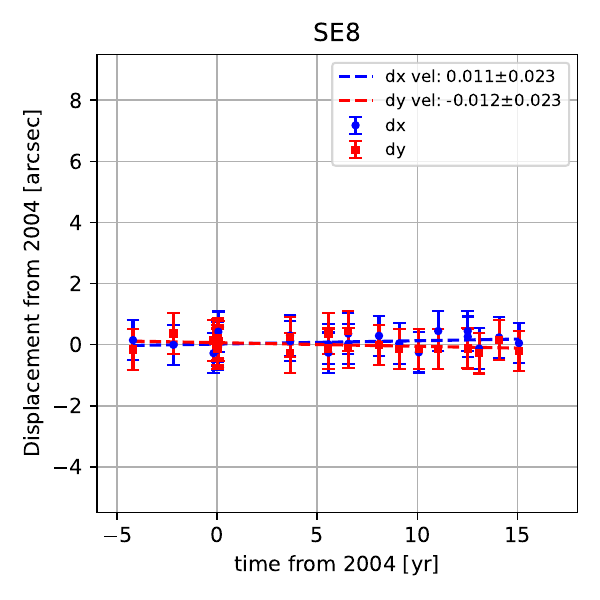}
\includegraphics[width=2cm]{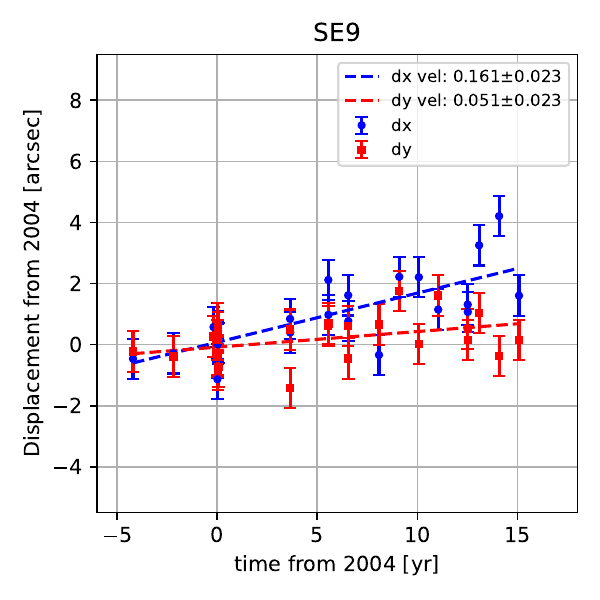}
\includegraphics[width=2cm]{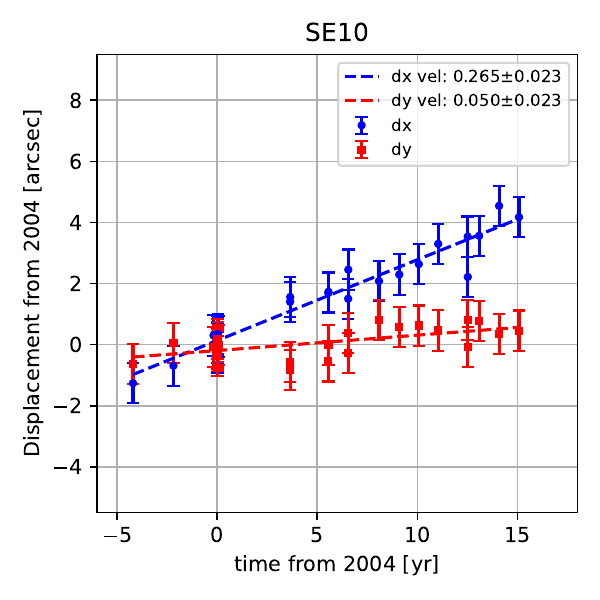}
\includegraphics[width=2cm]{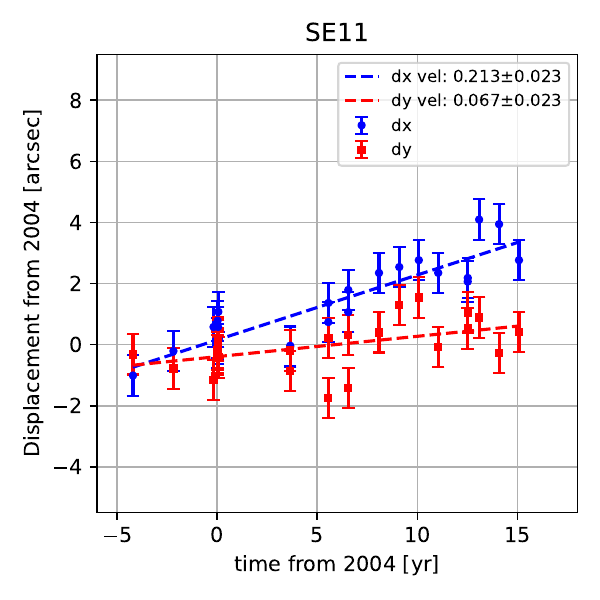}
\includegraphics[width=2cm]{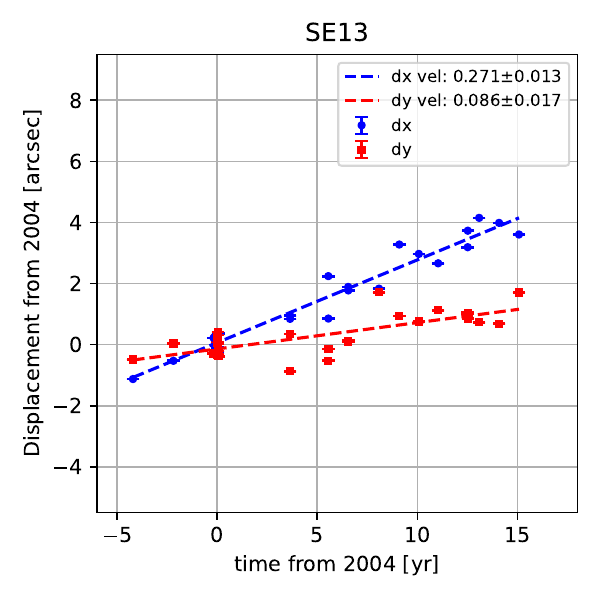}
\includegraphics[width=2cm]{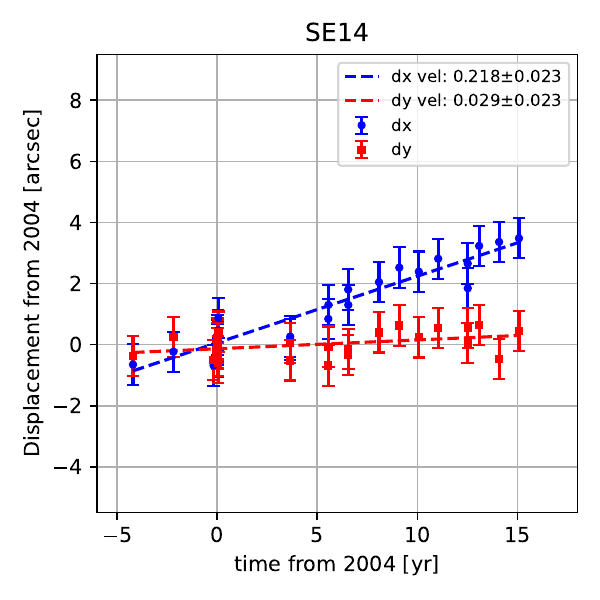}
\includegraphics[width=2cm]{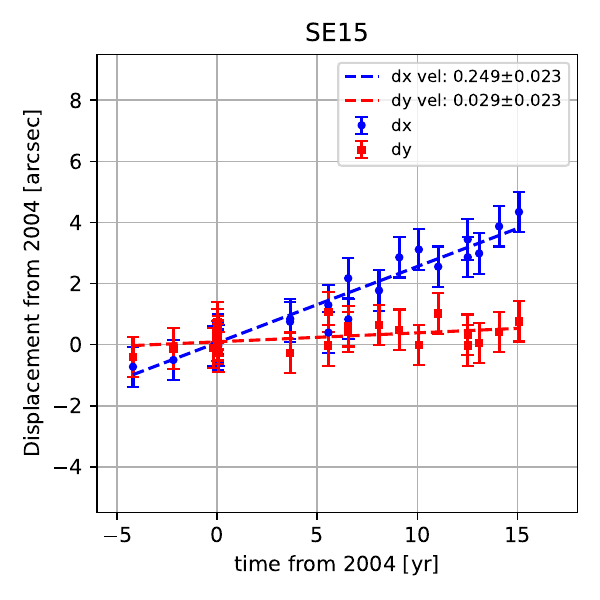}
\includegraphics[width=2cm]{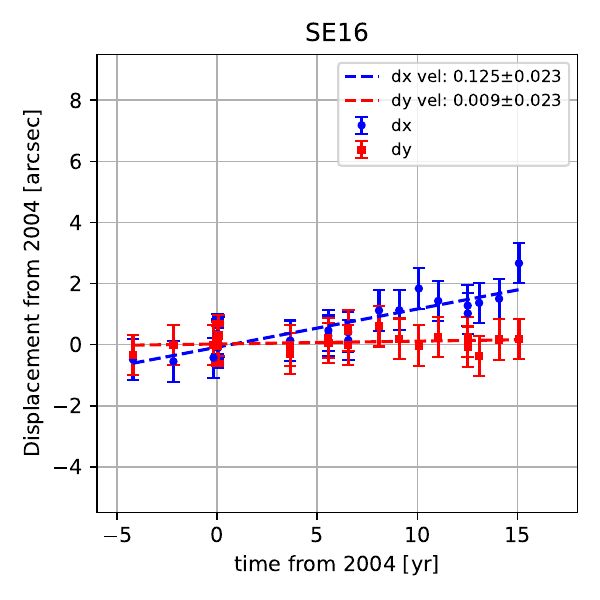}
\includegraphics[width=2cm]{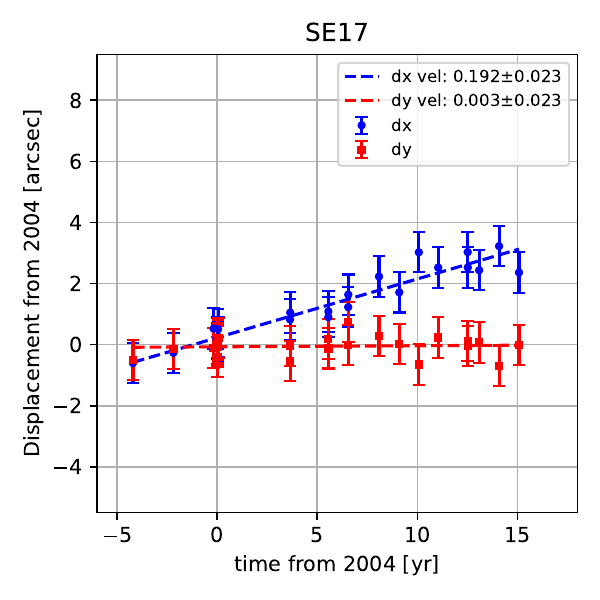}
\includegraphics[width=2cm]{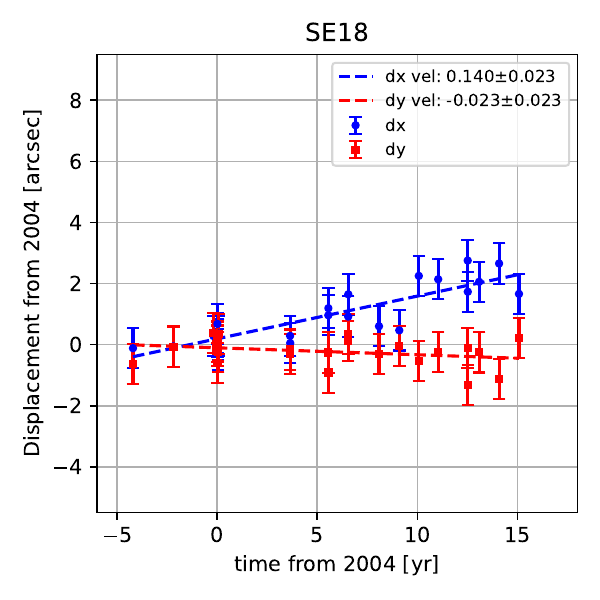}
\includegraphics[width=2cm]{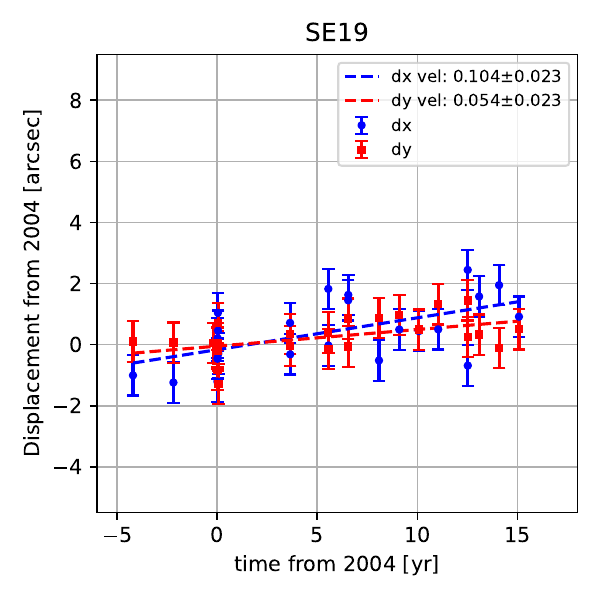}
\includegraphics[width=2cm]{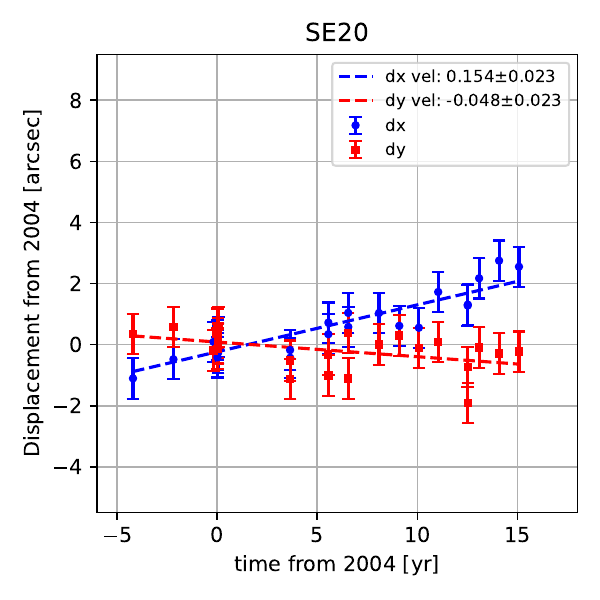}
\includegraphics[width=2cm]{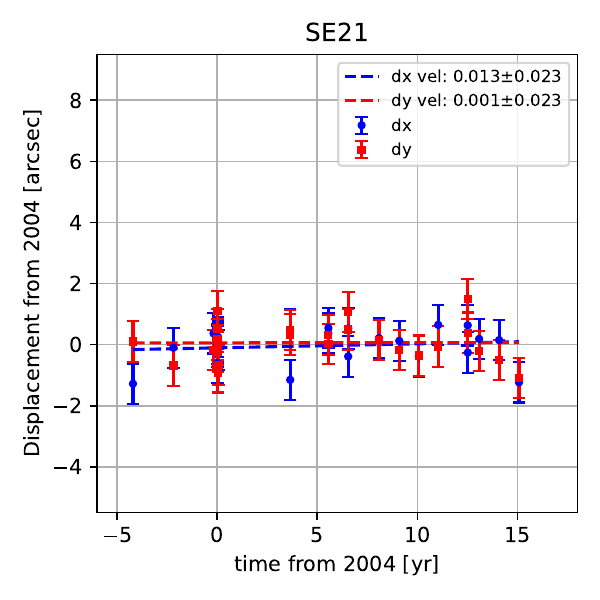}
\includegraphics[width=2cm]{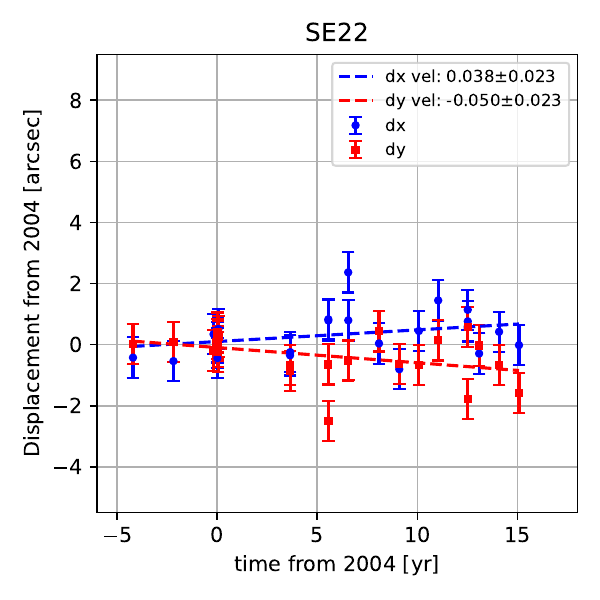}
\includegraphics[width=2cm]{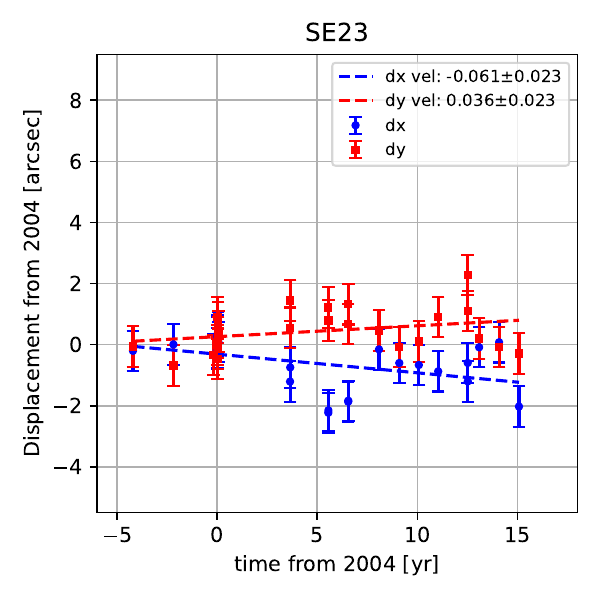}
\includegraphics[width=2cm]{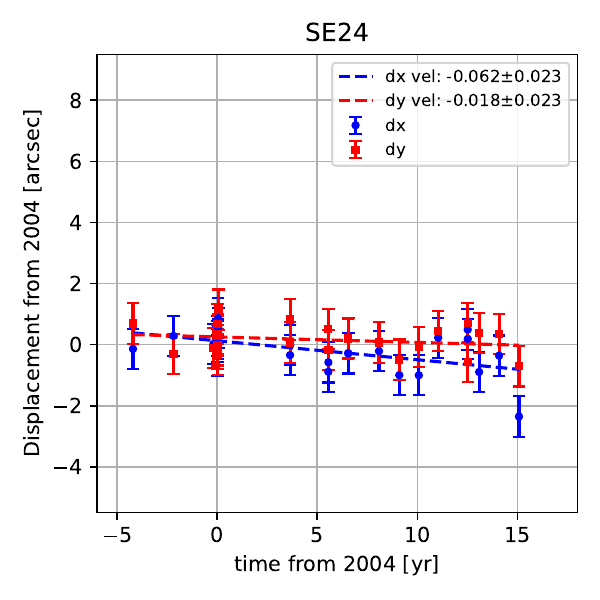}
\includegraphics[width=2cm]{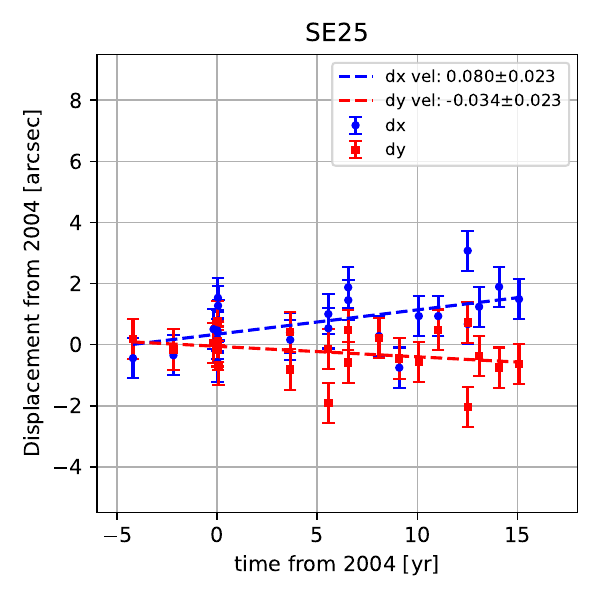}
\includegraphics[width=2cm]{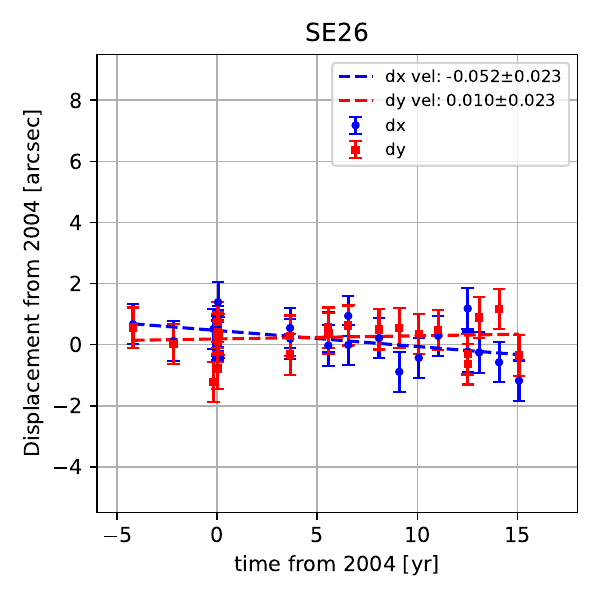}
\includegraphics[width=2cm]{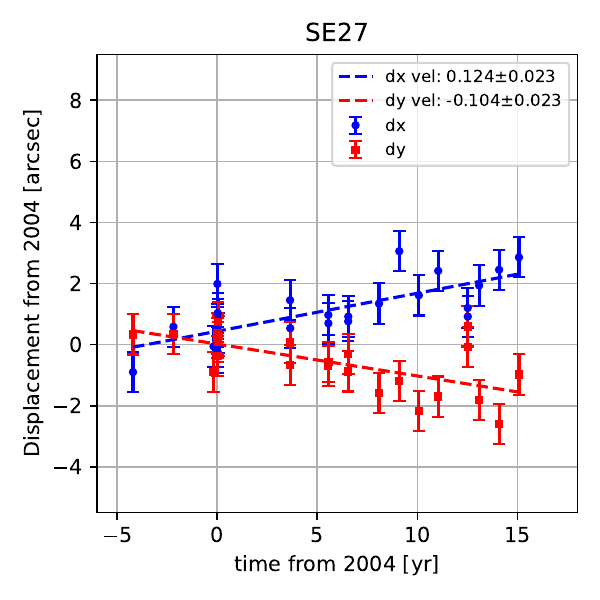}
\includegraphics[width=2cm]{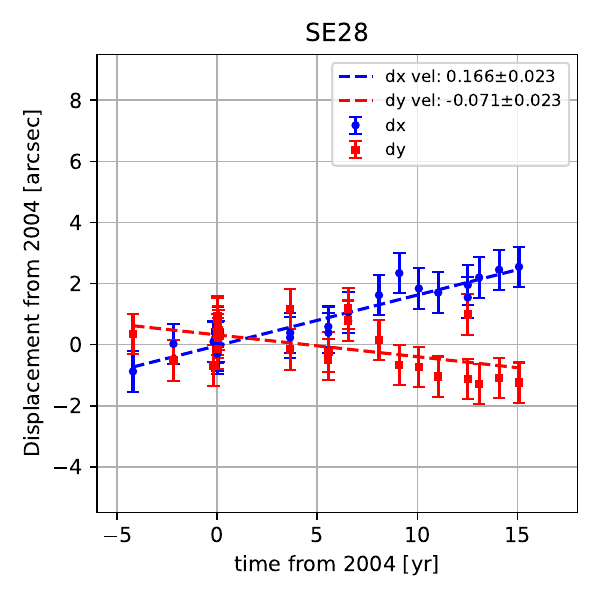}
\includegraphics[width=2cm]{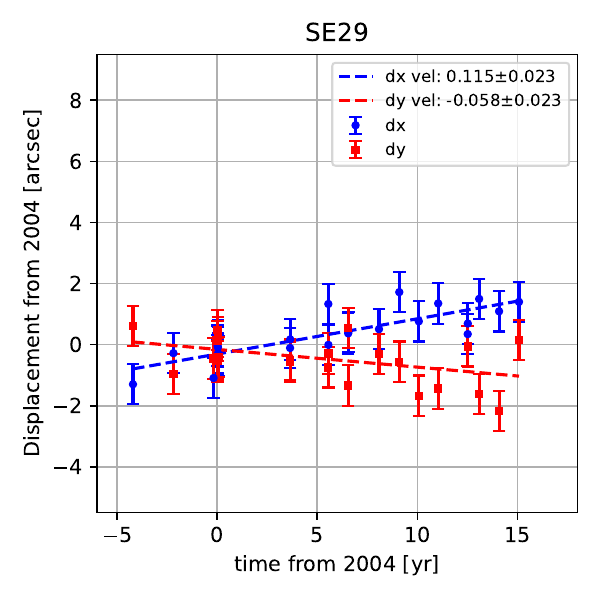}
\includegraphics[width=2cm]{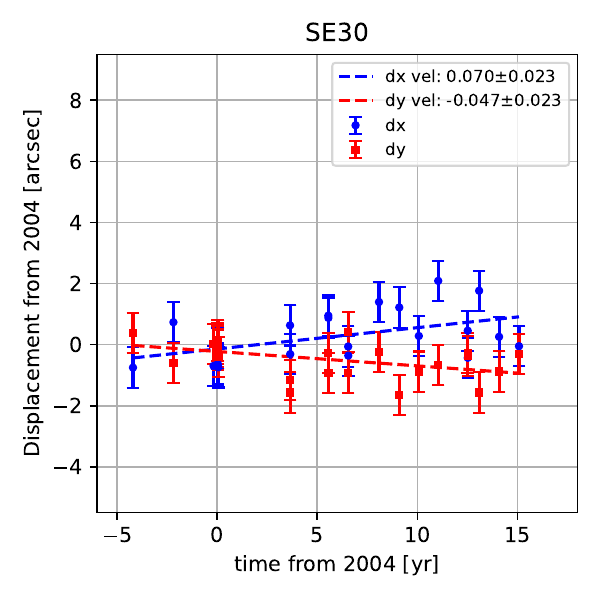}
\includegraphics[width=2cm]{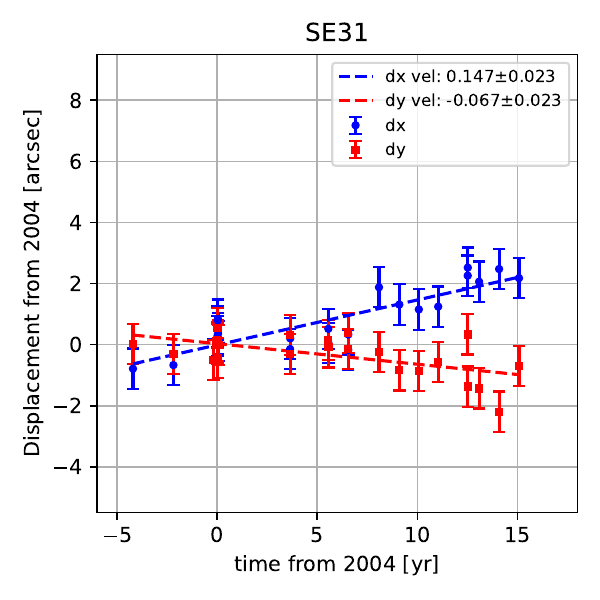}
\includegraphics[width=2cm]{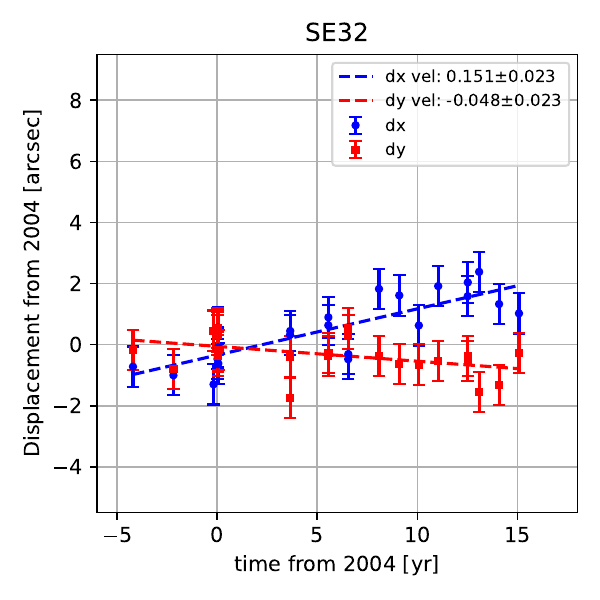}
\includegraphics[width=2cm]{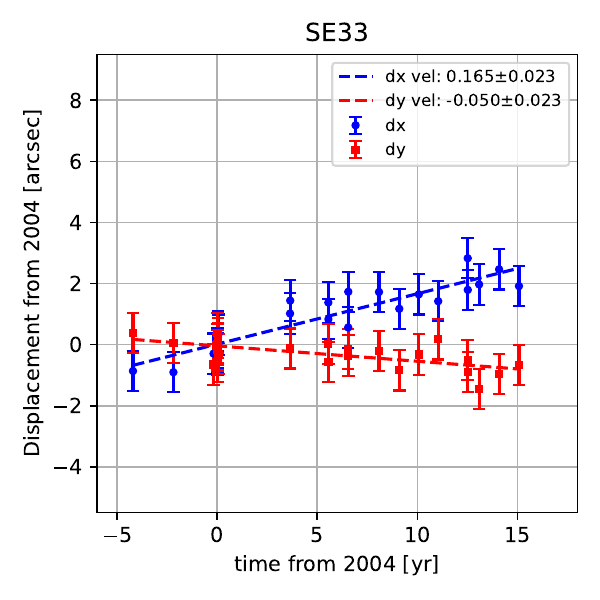}
\includegraphics[width=2cm]{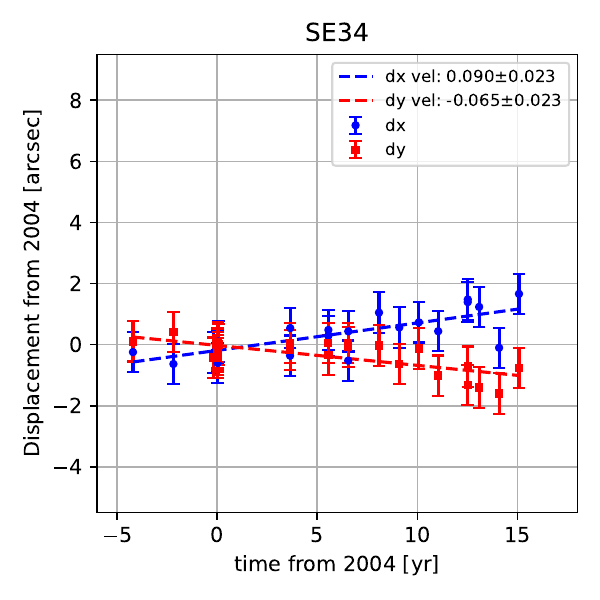}
\includegraphics[width=2cm]{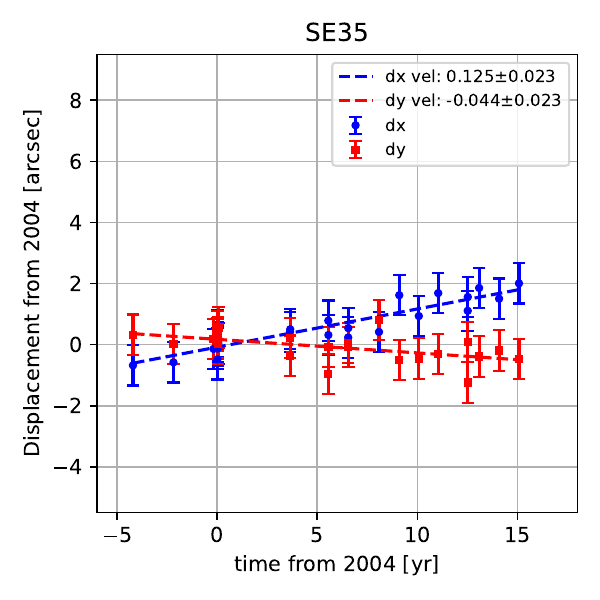}
\includegraphics[width=2cm]{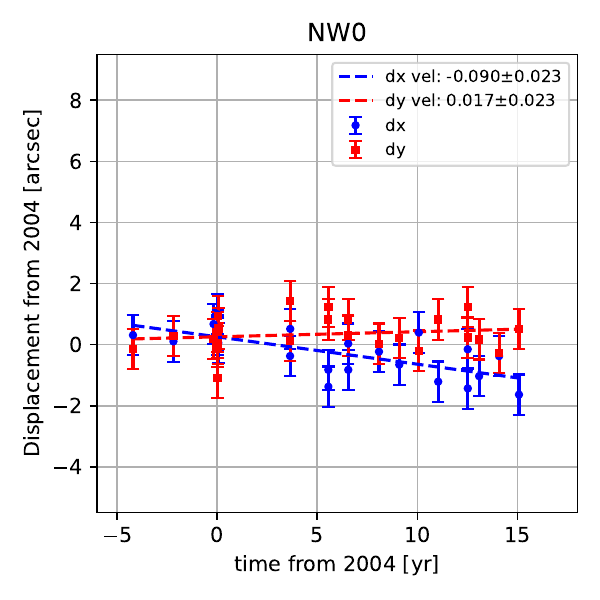}
\includegraphics[width=2cm]{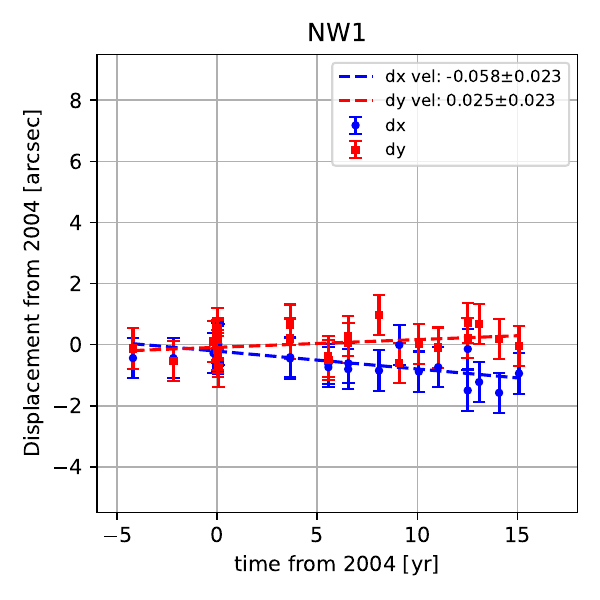}
\includegraphics[width=2cm]{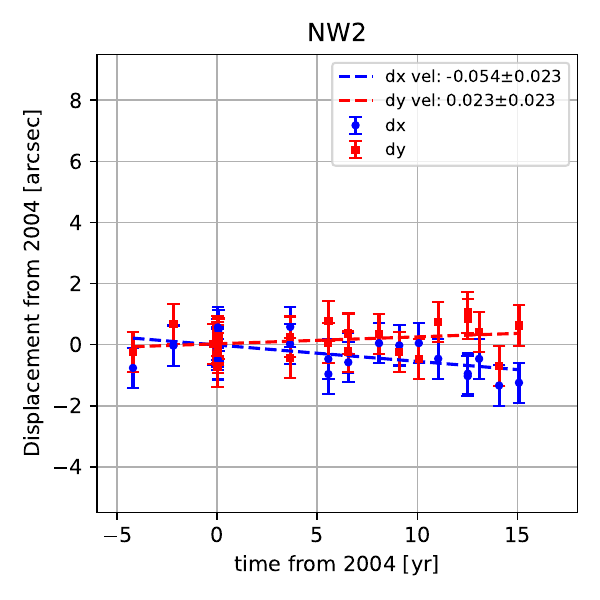}
\includegraphics[width=2cm]{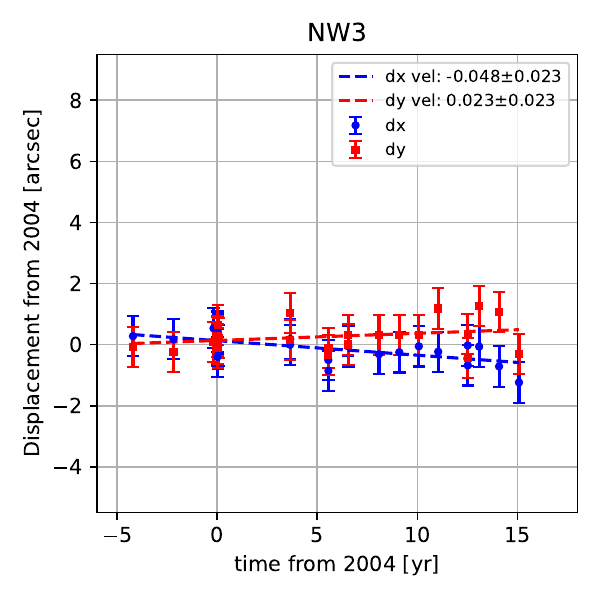}
\includegraphics[width=2cm]{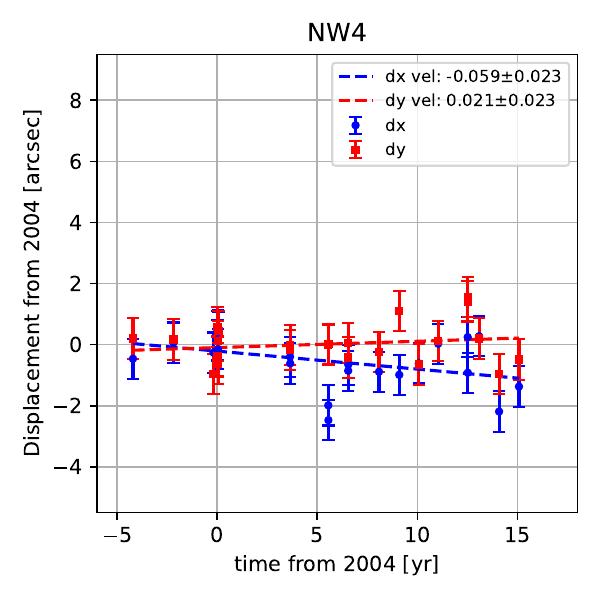}
\includegraphics[width=2cm]{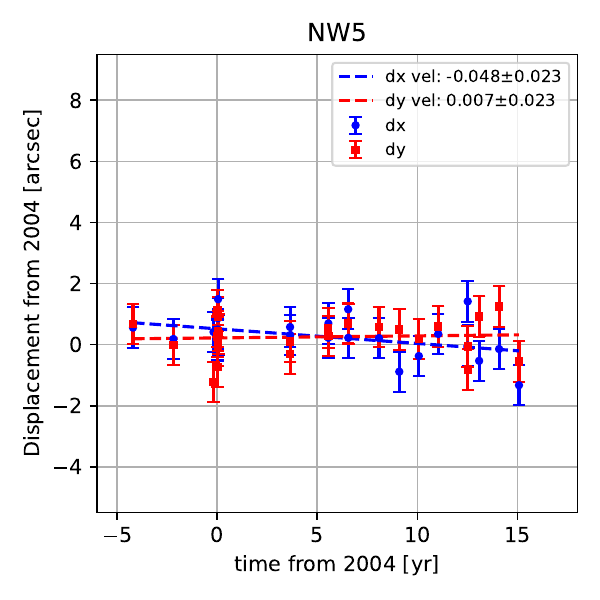}
\includegraphics[width=2cm]{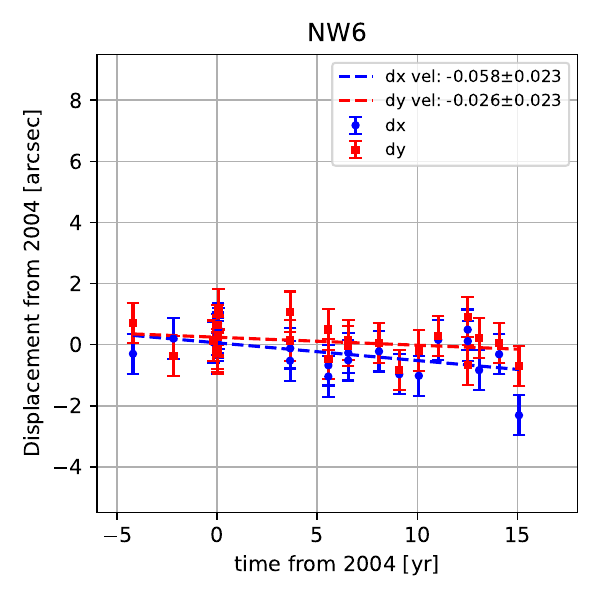}
\includegraphics[width=2cm]{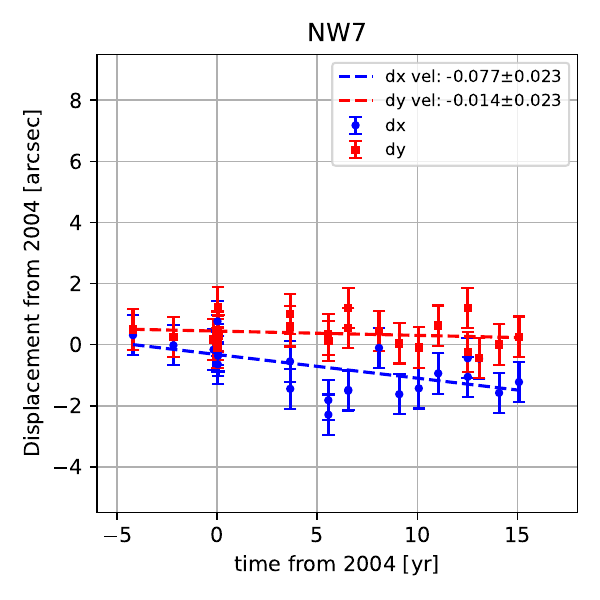}
\includegraphics[width=2cm]{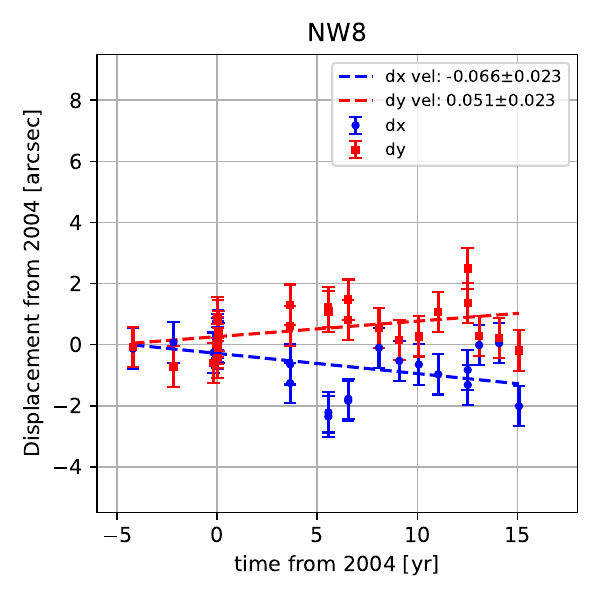}
\includegraphics[width=2cm]{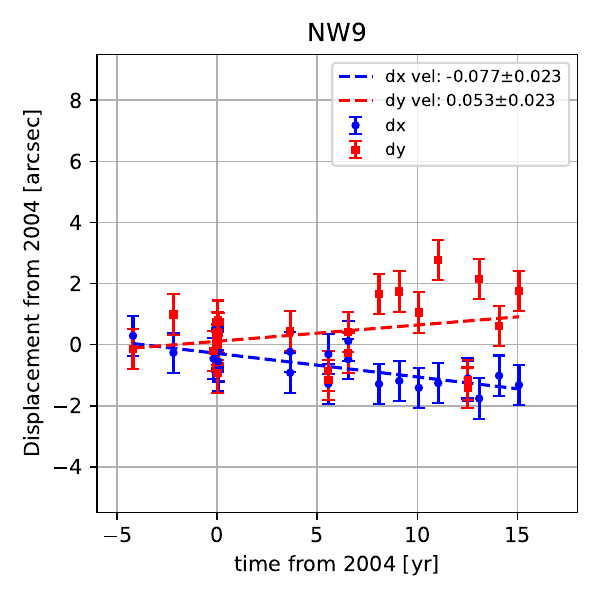}
\includegraphics[width=2cm]{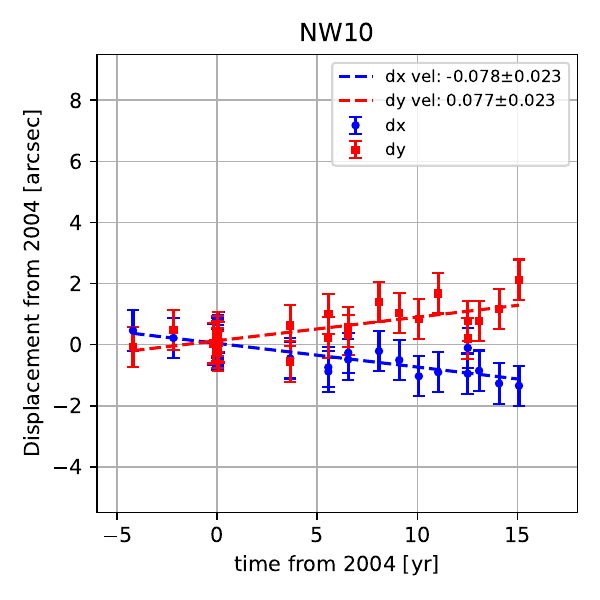}
\includegraphics[width=2cm]{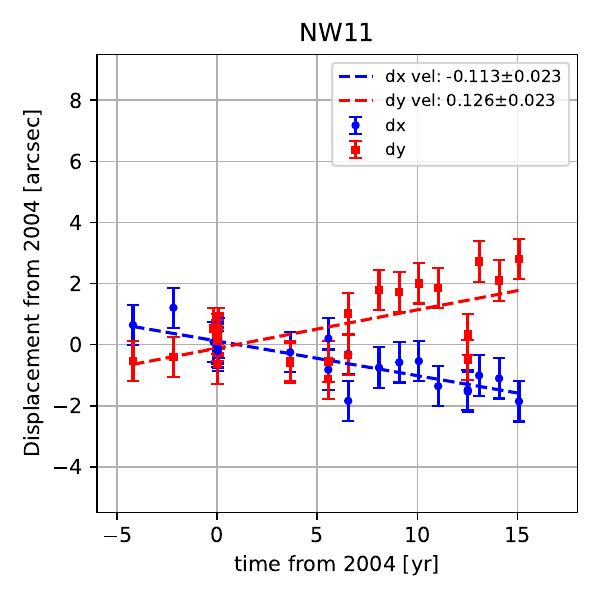}
\includegraphics[width=2cm]{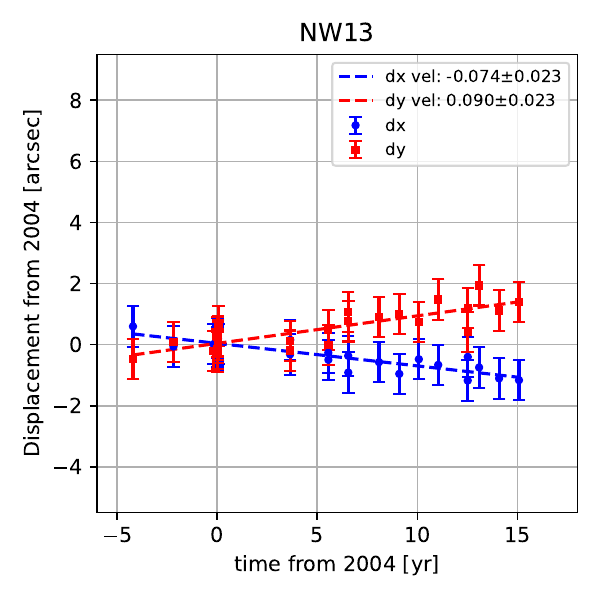}
\includegraphics[width=2cm]{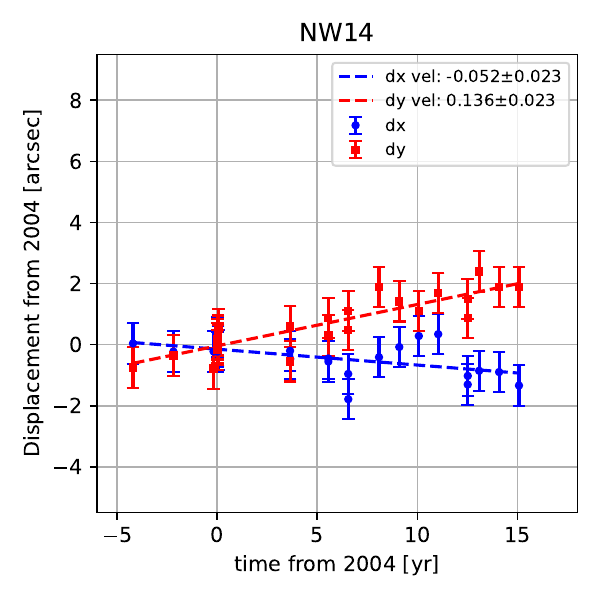}
\includegraphics[width=2cm]{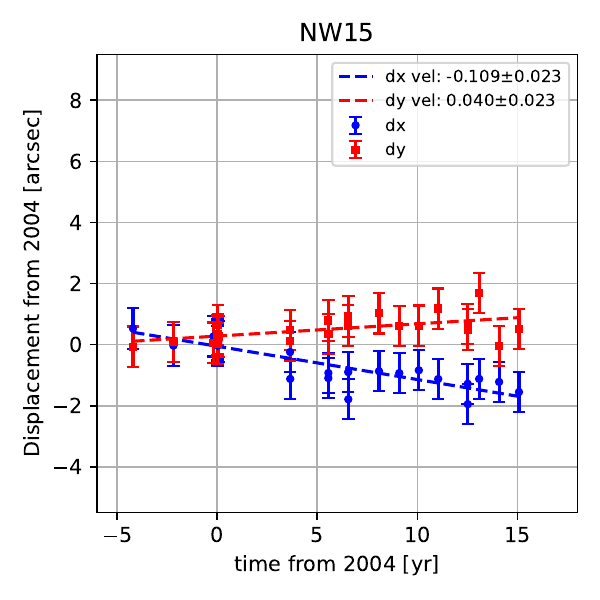}
\includegraphics[width=2cm]{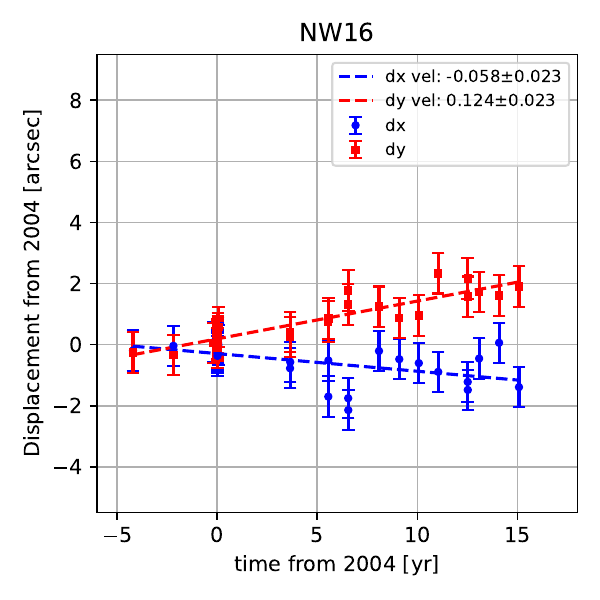}
\includegraphics[width=2cm]{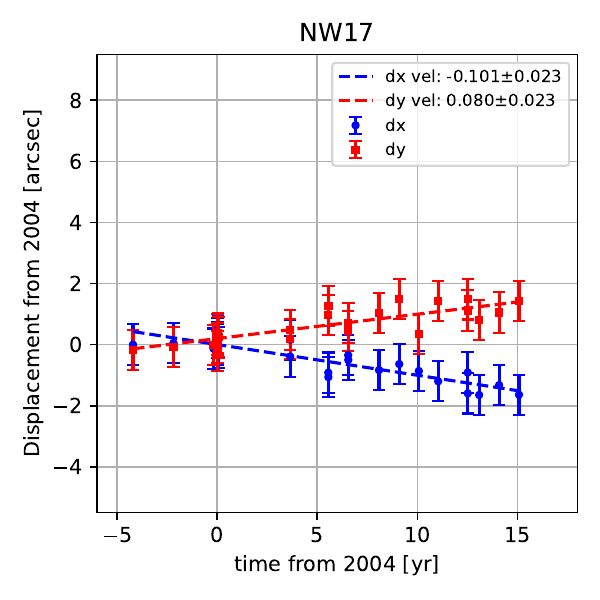}
\includegraphics[width=2cm]{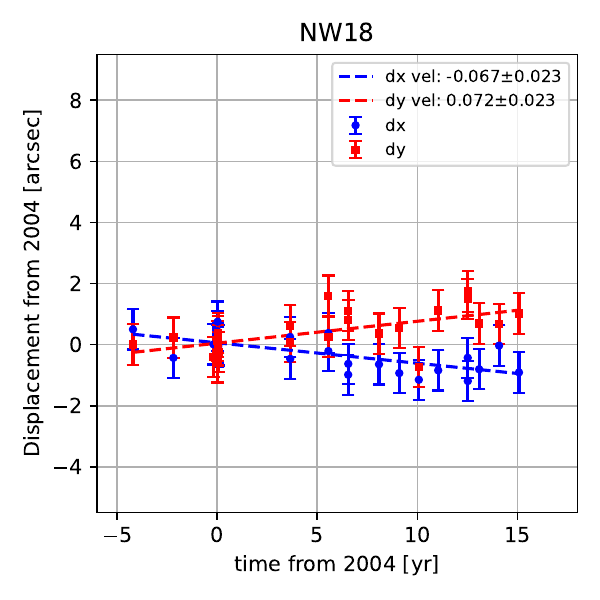}
\includegraphics[width=2cm]{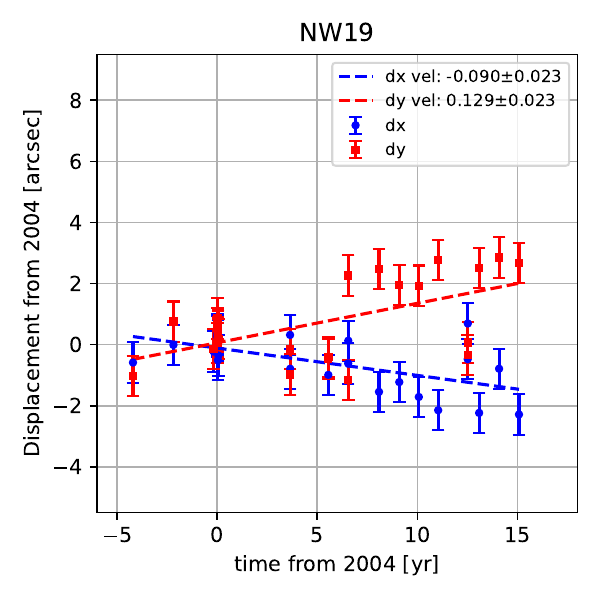}
\includegraphics[width=2cm]{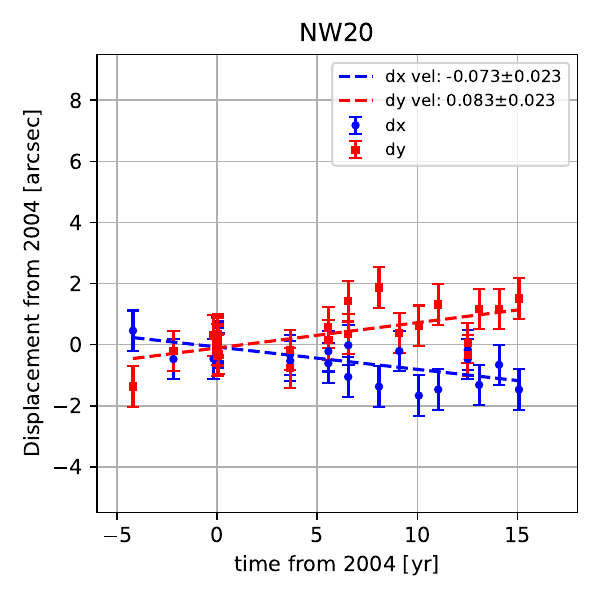}
\includegraphics[width=2cm]{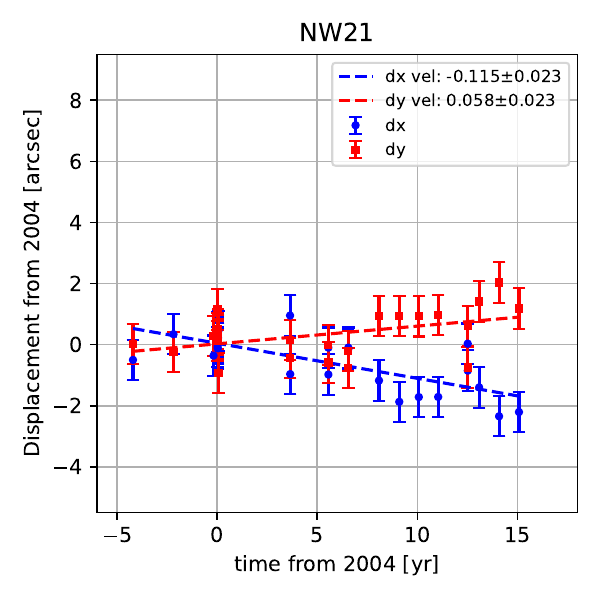}
\includegraphics[width=2cm]{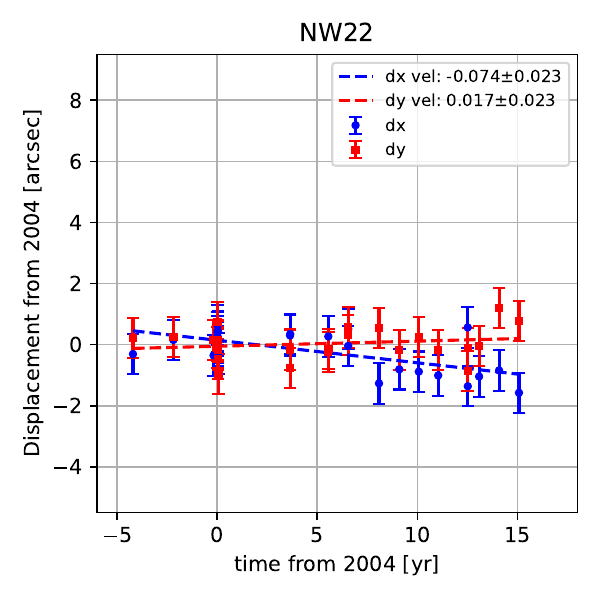}
\includegraphics[width=2cm]{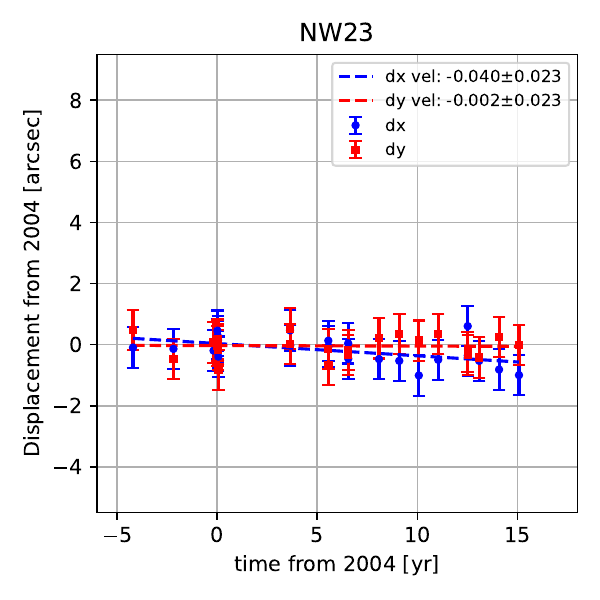}
\includegraphics[width=2cm]{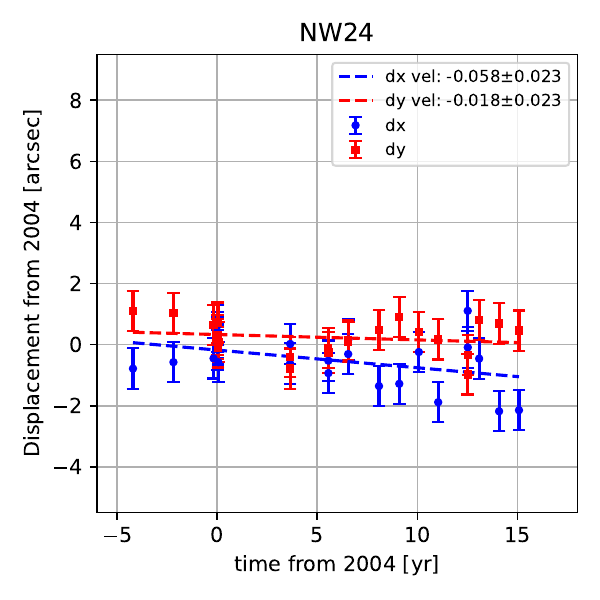}
\includegraphics[width=2cm]{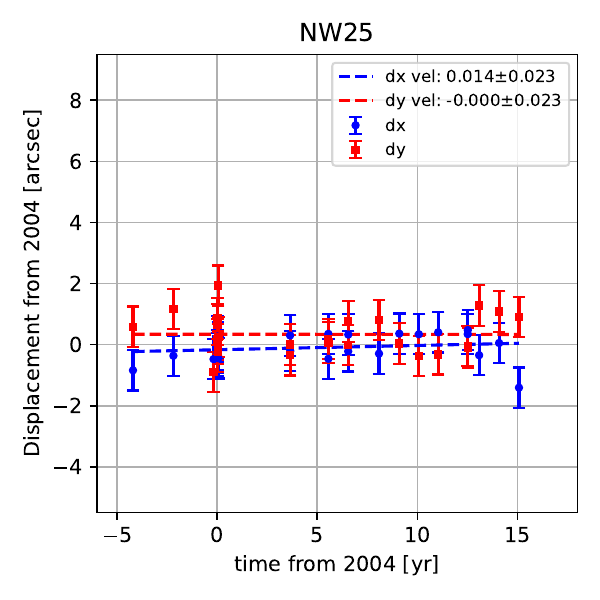}
\includegraphics[width=2cm]{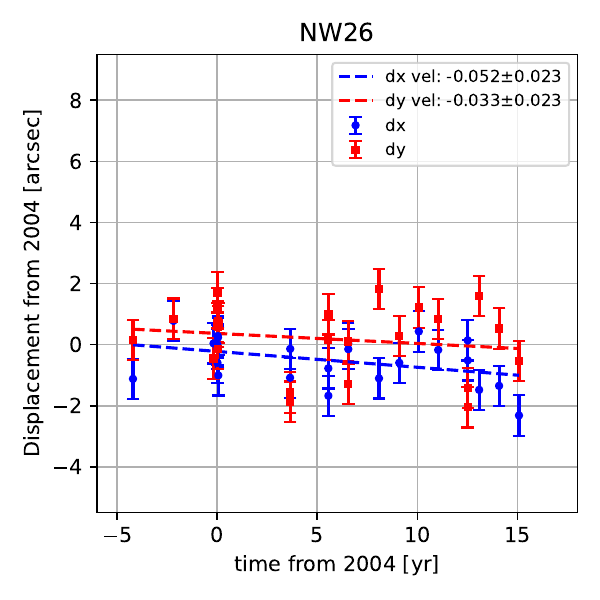}
\includegraphics[width=2cm]{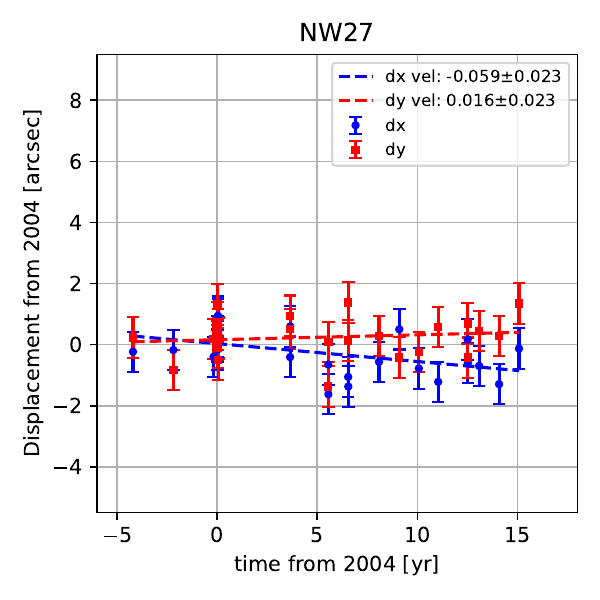}
\includegraphics[width=2cm]{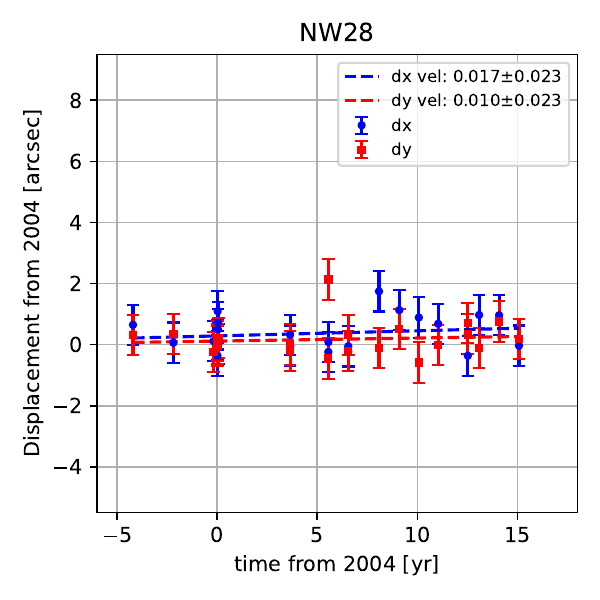}
\includegraphics[width=2cm]{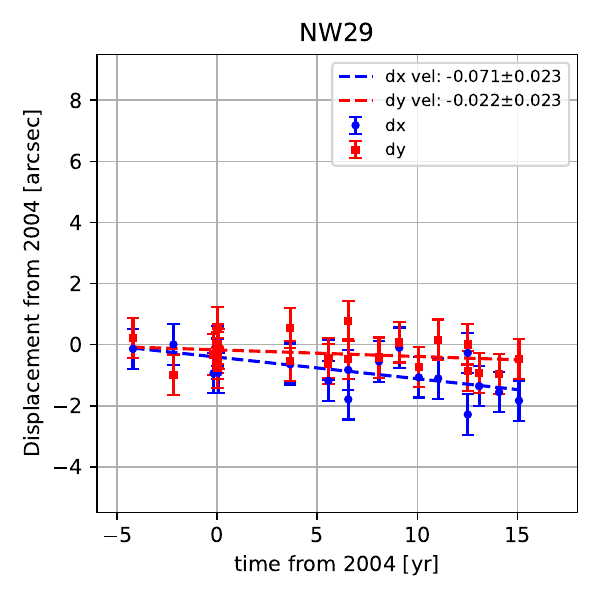}
\includegraphics[width=2cm]{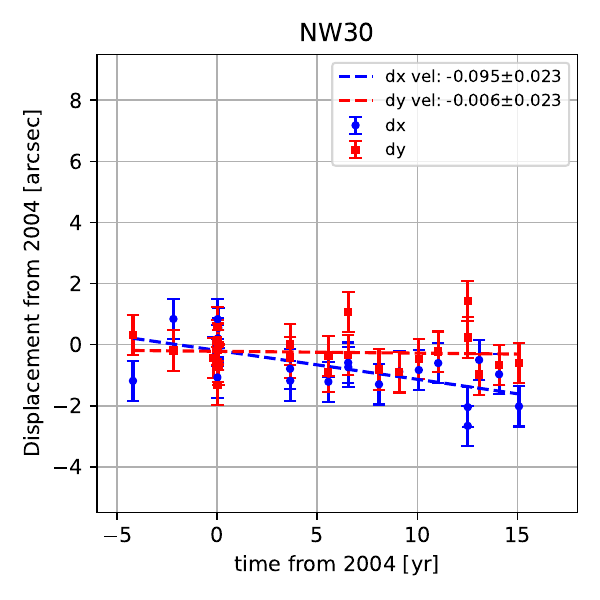}
\includegraphics[width=2cm]{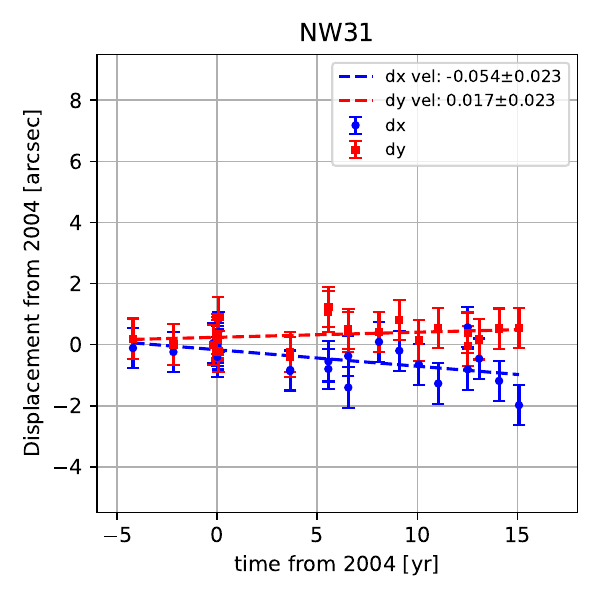}
\includegraphics[width=2cm]{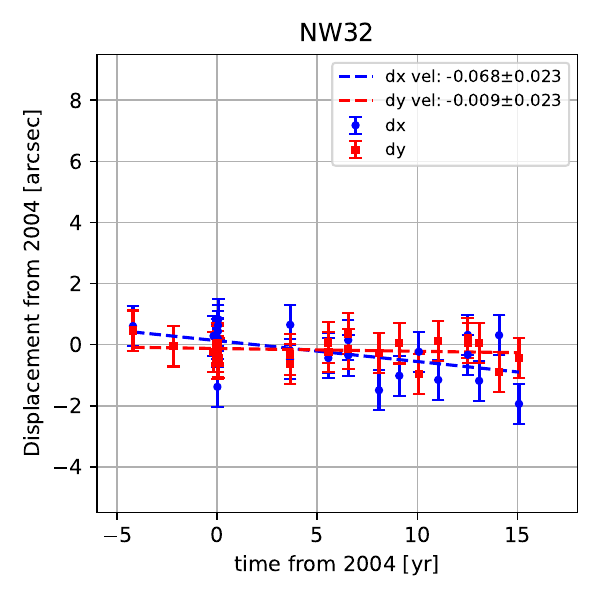}
\includegraphics[width=2cm]{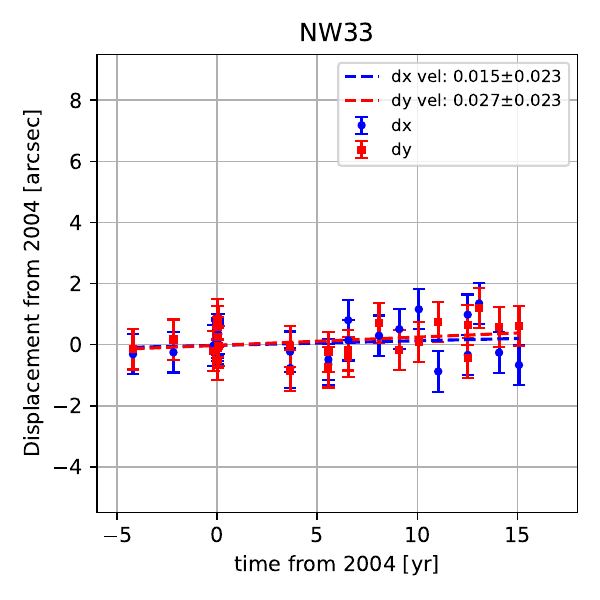}
\includegraphics[width=2cm]{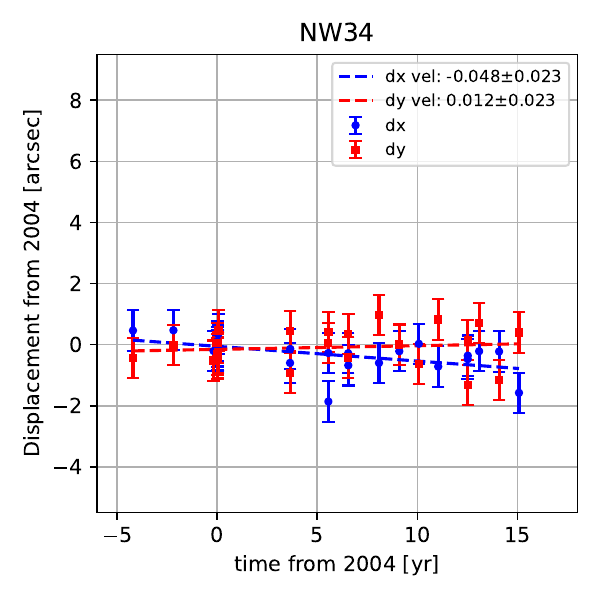}
\includegraphics[width=2cm]{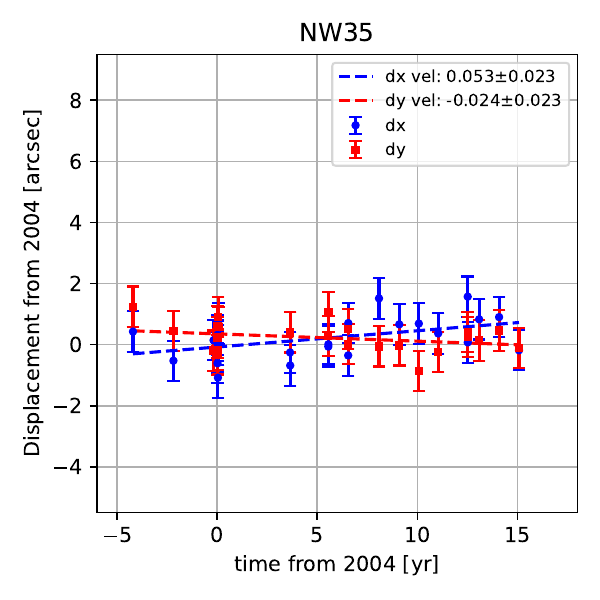}
\end{center}
\caption{$dt$ vs. $dx$ (blue) and $dy$ (red) for each pixel.
Dashed lines represent the best-fit line of the displacement.
{Alt text: Line graphs showing dt vs. dx and dy for each pixel.}}\label{fig:dx-dy}
\end{figure}

In this section, we summarize the proper motion measurement results for each pixel.
The results are shown in Figure~\ref{fig:dx-dy}.

\section*{Appendix B: Fe-K proper motion measurements using radial-expansion, multi-epoch fitting}
\label{appendix_jacco}

The proper motion measurements  reported in the main text are based on the 
optical flow algorithm by \citet{farneback2003}. It offers results by a well-established algorithm, but  it has a drawback that it does not take into
account the Poissonian noise of the  astrophysical images used \citep[for example]{sakai2024}.
Instead the statistical uncertainties are estimated from the intrinsic scatter of the proper motion solutions comparing the many individual base-lines used.

In order to check the validity of these results presented, we offer here proper motion measurements using the method
and code used in \citet{vink2022}.  It is based on the maximum likelihood method for Poissonian statistics, the so-called  C-statistic \citep{cash1979}.
Instead of proper motion measurements per time interval, the method uses multi-epoch fitting, producing
one C-statistic per region. The combined data from the
deep 2004 Chandra observations  (OBSID=4634--4639; \cite{hwang2004}) serve as the model. 
Per selected region and per epoch this model image is expanded with a factor $f=a \Delta t$,
with $a$ the expansion factor to be measured,
and $\Delta t$ the time difference between the mean observation data for the 2004 Chandra data and the specific epoch used as comparison. 
For each region one obtains a best-fit $a$ based on combining the C-statistic of all individual bases lines together.
Labeling the region $j$ and the baseline $i$, the code determines $a_j$ by calculating $f_{ij} =a_{j}\Delta t_i$, by minimizing $\ln L_j= \sum_i L_{ij}$.

A problem for the X-ray proper motions for Cas A is that the Chandra field of view  does not  contain multiple point sources, which can be used
to correct aspect errors. 
There is only one point source, the stellar remnant of the explosion, which has its own proper motion, as inferred from its position southeast of the explosion center.
In order to correct for pointing accuracies we used the iterative procedure outlined in  \citet{vink2022}:
after determining for each region the expansion factor $a_j$, we run the code again, but now fixing $a_j$ and now determining the pixel offsets 
$\delta x_i, \delta y_i$ for each epoch/observation, by combining the C-statistic likelihood for all regions $j$: $\ln L_i = \sum_j L_{ij}$.
After determining $\Delta x_i, \Delta y_i$ we run the code again to determine for each region $a_j$. This iterative procedure was repeated several times
until the changes in  $\Delta x_i, \Delta y_i$ per iteration were less than 0.1 ACIS-S pixels (0.049\arcsec). As can be seen in Table~\ref{tab:epochs}
the corrections themselves are of the order of 0.5 pixels (0.47\arcsec).
The details of the method can be found in \citet{vink2022}. We did not use the same astrometric solution here, as we used the
\texttt{CIAO v4.17} version of \texttt{chandra\_repro} script to reprocess the data.

We would like to stress here that the method outlined in this appendix offers the advantage of a workable astrometric solution and proper-motion measurements
in which the proper motion measurements are calculated taking into account poissonian noise, but it has two disadvantages:
(1) it only measures the radial expansion with respect to the Cas A explosion center, for which we used the measurement by \citet{thorstensen2001};
(2) the method does not allow the subtraction of the continuum contributions to the 6.5--6.75 keV Fe K band used.

\begin{figure*}
  \includegraphics[height=0.33\textwidth,trim=0 0 0 0,clip=true]{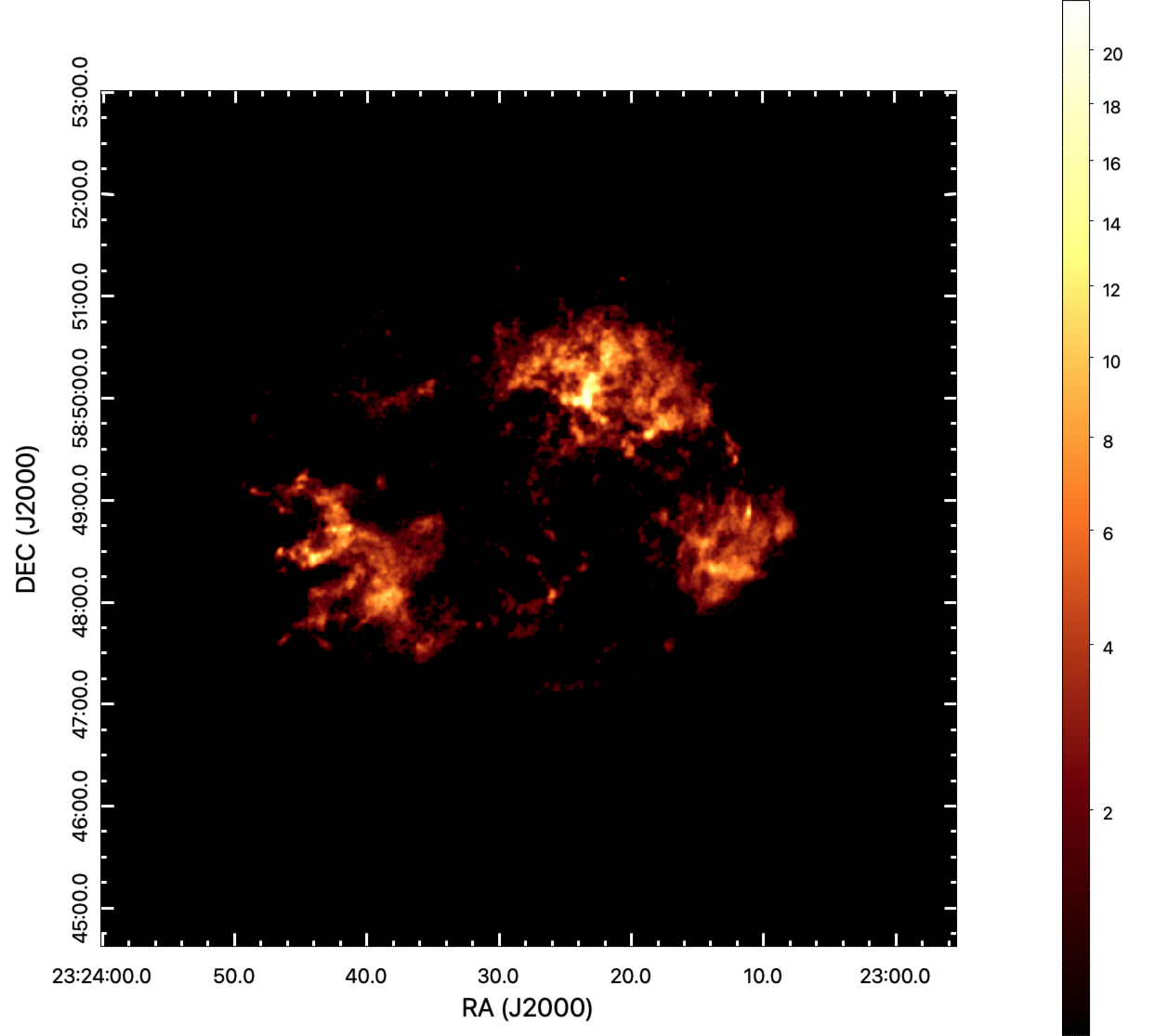}
  \includegraphics[height=0.33\textwidth,trim=0 0 140 0,clip=true]{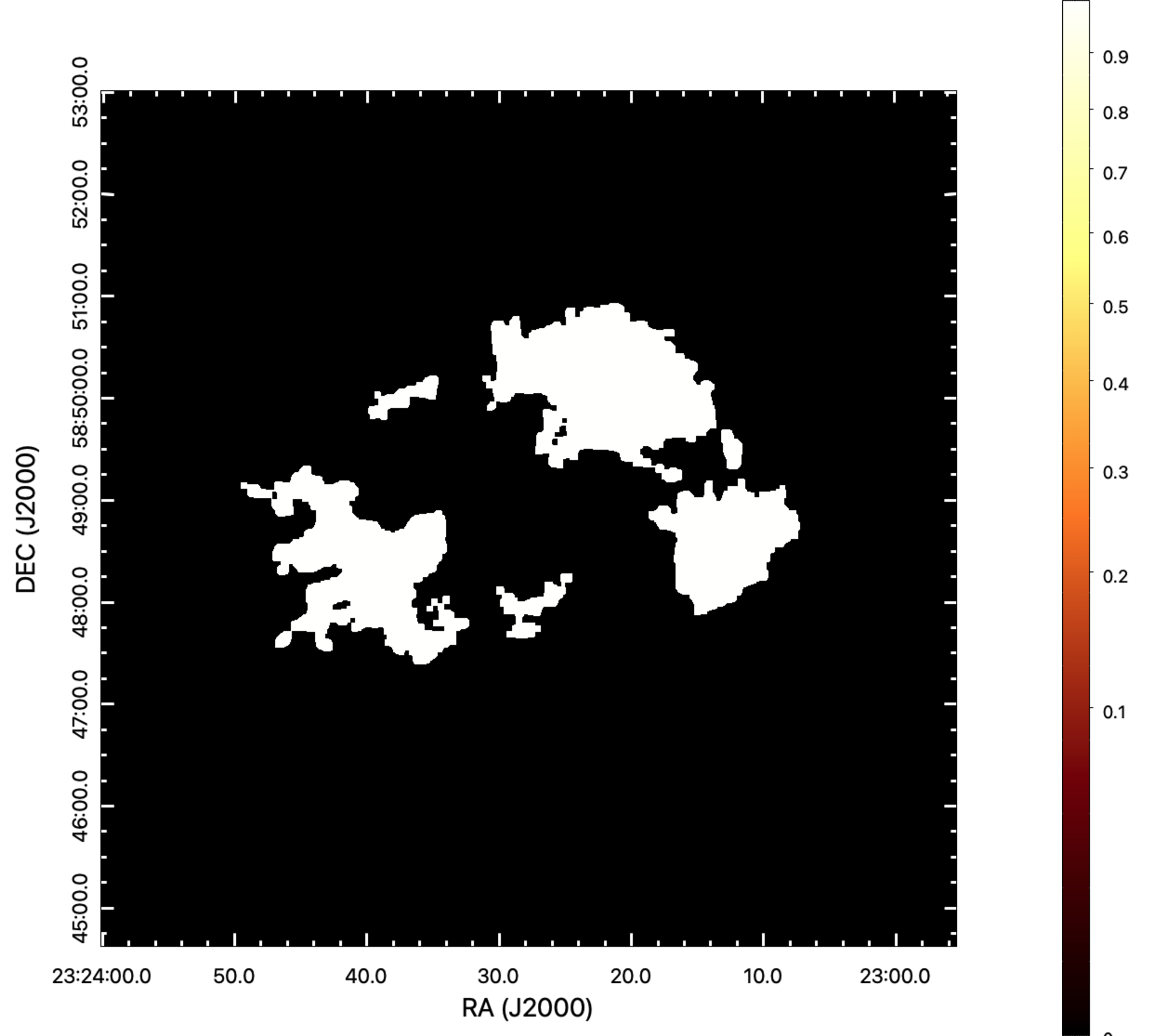}
  \includegraphics[height=0.33\textwidth,trim=0 0 140 0,clip=true]{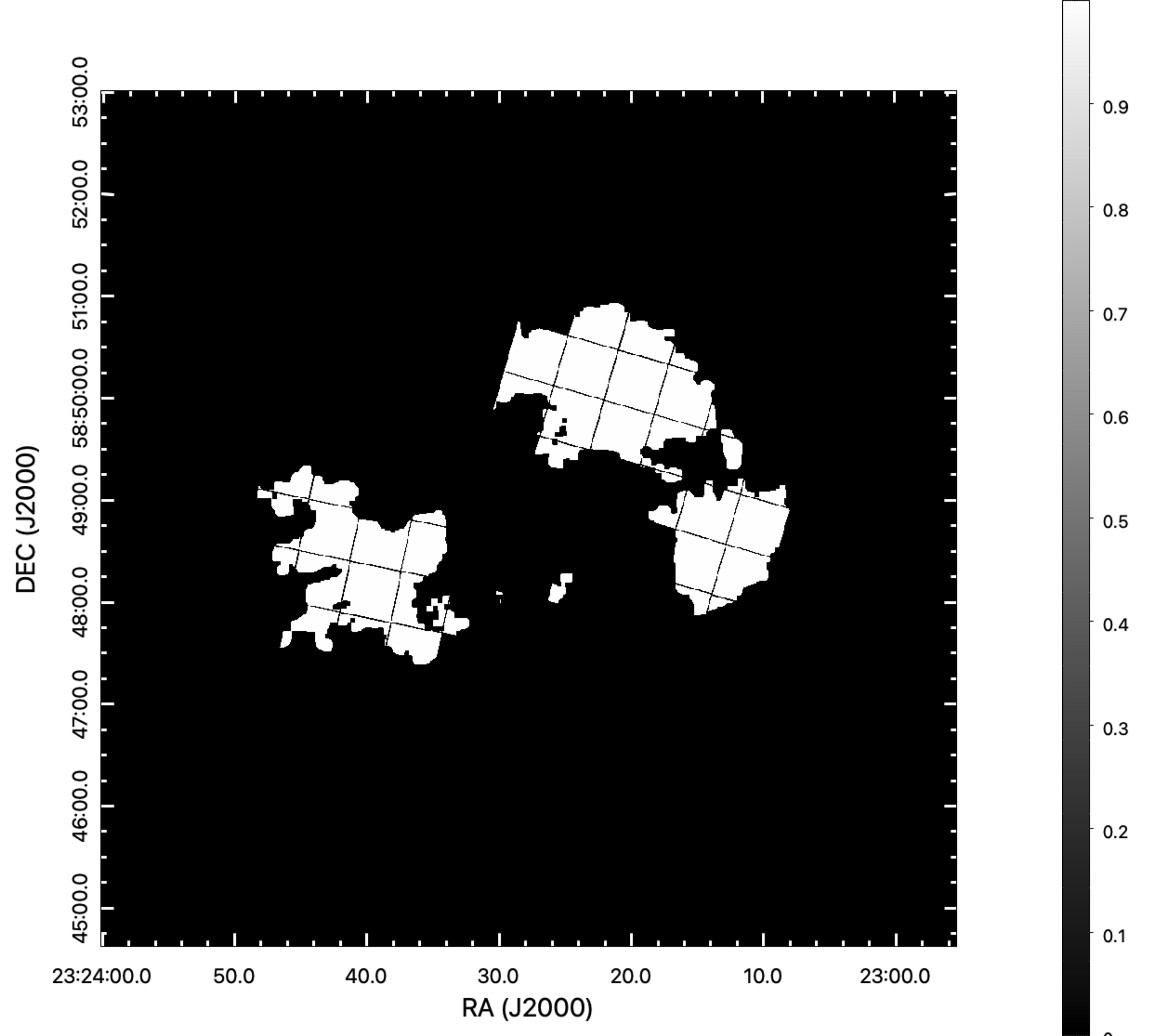}
  \caption{Left: The smoothed and clipped Fe K map of Cas A, based on the 2004 VLP data. This image forms the basis for making an Fe K bright
    mask.
    Center: the mask derived from the Fe K map. Right: The mask for the Fe K map with the grid of the available XRISM/Resolve pixels
    for the NW and SE pointings.
    {Alt text: Fe K maps with Chandra.}
  \label{fig:fe_masks}}
\end{figure*}

\begin{figure*}
\centerline{
\includegraphics[width=0.45\textwidth]{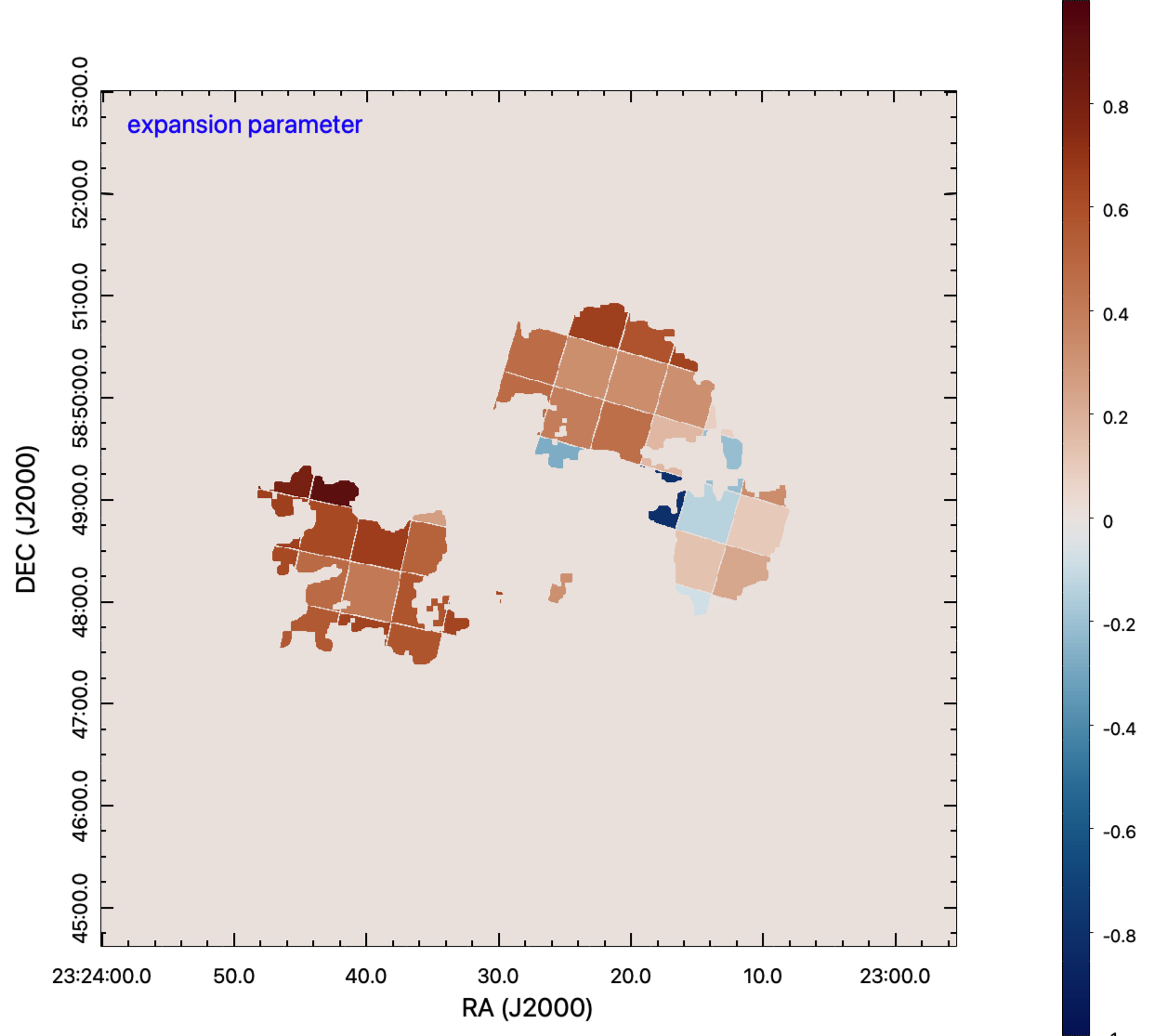}
\includegraphics[width=0.45\textwidth]{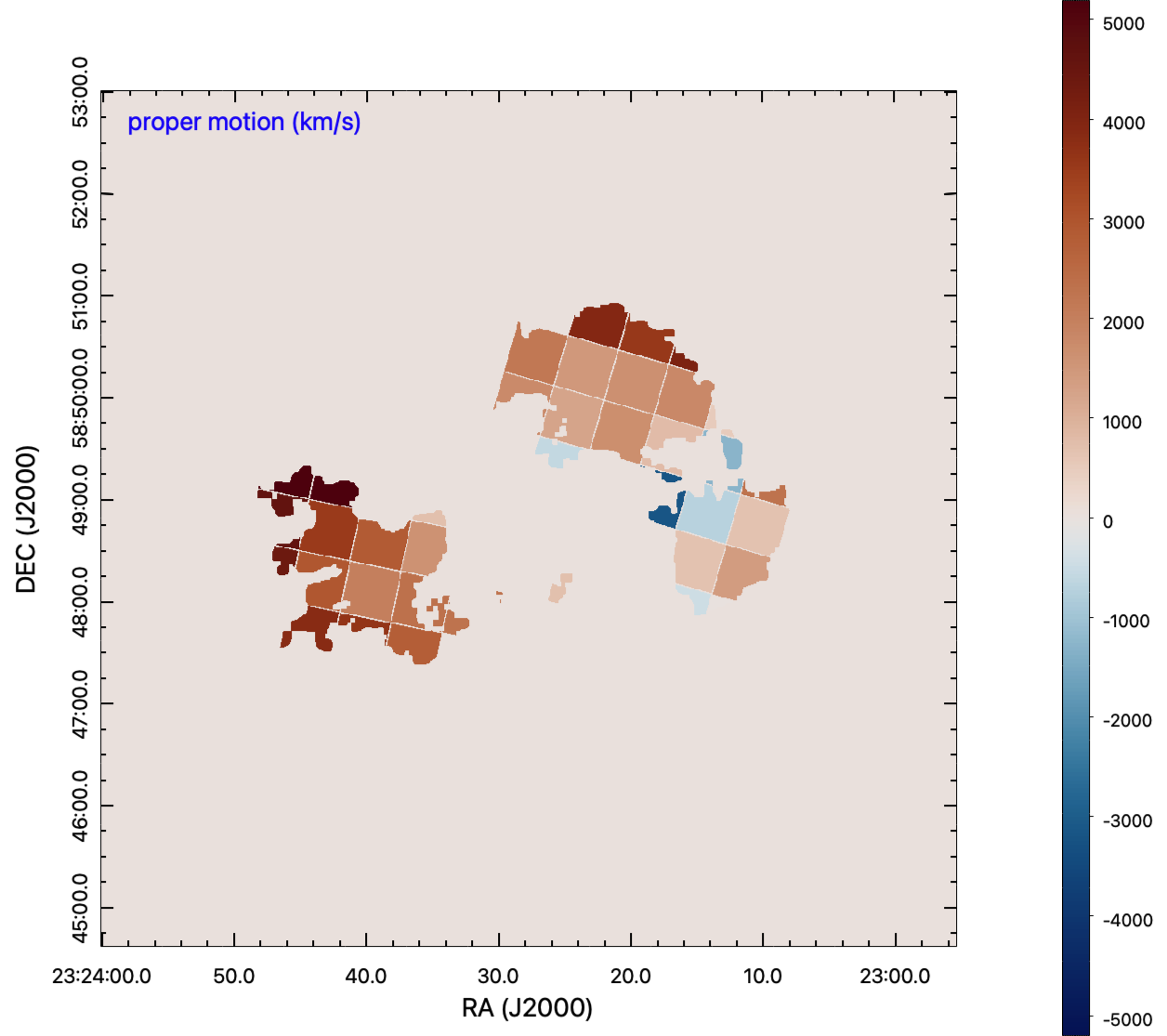}
}
\caption{\label{fig:fek_exp}
Results of the radial expansion measurements. Left: the expansion parameter, defined as $m\equiv t_{\rm Cas A}/\tau_{\rm exp} \propto a_j$, with $t_{\rm Cas A}=332$ the approximate age
of Cas A in 2004 \citep{thorstensen2001}. Right: the measured radial proper motions $v_{\rm prop}$ in km\,s$^{-1}$.
    {Alt text: Expansion measurement maps of Fe K.}
}
\end{figure*}

To mitigate the second disadvantage, we only measured proper motions for those regions for which
 the number of counts in the combined 2004 Chandra Fe-K image
  exceeds a threshold of 1 count per pixel after smoothing with a Welch filter with $7\times 7$ pixels. We used this image to obtain a cleaned-up version using pixel dilation and erosion to
  remove isolated pixels above the threshold, and 
  merge smaller islands ``archipels" into larger structures. As we are interested in combining
with the radial proper motion measurements by XRISM/Resolve, we finally made masks for each XRISM/Resolve pixel.
The three different steps to obtain the final extraction regions for proper motions measurements are shown in Fig.~\ref{fig:fe_masks}.
The results of the expansion measurements are 
visualized in Fig.~\ref{fig:fek_exp}.

Noteworthy are the very fast expansion of the eastern most iron protrusion (below the jet opening), where the expansion parameter
is 0.78--0.82 and $v_{\rm prop}\approx 4700$--$5600$~km~s$^{-1}$, and the negative  (inward) proper motions in the west, which mimic
the inward velocities measured in the continuum band \citep{sato2018,vink2022,sakai2024}.

\begin{table}
\centering
\caption{Summary of the observations used and astrometric corrections.\label{tab:epochs}}
\begin{tabular}{lrccrcc}\hline\hline\noalign{\smallskip}
Epoch & ObsID &  MJD & $\Delta t$ & Exposure & $\Delta x$ & $\Delta y$\\
 & &  &  {[yr]} & {[ks]} & [pixel] & [pixel]\\\noalign{\smallskip}\hline\noalign{\smallskip}
"Model" & 4634--4639 & 53121.3 & 0 & 845.6 &\\
Epoch 1 & 114 & 51573.8 & -4.237 & 50.6 &  0.10    &        0.66 \\
Epoch 2 & 1952 & 52311.6 & -2.217 & 50.3 &  0.10    &        0.40 \\
Epoch 3 & 9117 & 54440.1 & 3.611 & 25.2 &  -0.18    &        0.29 \\
Epoch 4 & 9773 & 54442.7 & 3.618 & 25.2 & -0.45     &       0.44  \\
Epoch 5 & 10935 & 55138.1 & 5.522 & 23.6 & -0.62     &       0.37\\
Epoch 6 & 12020 & 55500.4 & 6.513 & 32.7 & -0.81    &       -0.43 \\
Epoch 7 & 10936 & 55139.1 & 5.524 & 22.7 &  -0.76  &          0.22 \\
Epoch 8 &  13177& 55502.2 & 6.519 & 17.5 &  -0.89 &           -0.70 \\
Epoch 9 & 14229 & 56062.7 & 8.053 & 49.8 &  -0.85  &         -1.04 \\
Epoch 10 & 14480 & 56432.9 & 9.067 & 49.4 & -0.08  &         -0.29  \\
Epoch 11 & 14481 & 56789.4 & 10.043 & 50.1 &  -0.64 &           0.13   \\
Epoch 12 & 14482 & 57142.8 & 11.010 & 50.1 &  -0.66 &          -0.80  \\
Epoch 13 & 18344 & 57682.9 & 12.489 & 26.1 & -0.47  &          0.40  \\
Epoch 14 &  19903 & 57890.0 & 13.056 & 50.2 &  -0.56  &         -0.43  \\
Epoch 15 &  19604 & 58254.0 & 14.052 & 50.1 &   -0.02  &          0.35    \\
Epoch 16 &  19606 & 57681.3 & 12.485 & 25.0 &  -0.37  &          0.17 \\
Epoch 17 &  19606 & 58616.8 & 15.046 & 50.1 &   -1.06 &          -0.35   \\
\noalign{\smallskip}\hline
\end{tabular}
\end{table}

\begin{figure}
\includegraphics[width=0.5\textwidth]{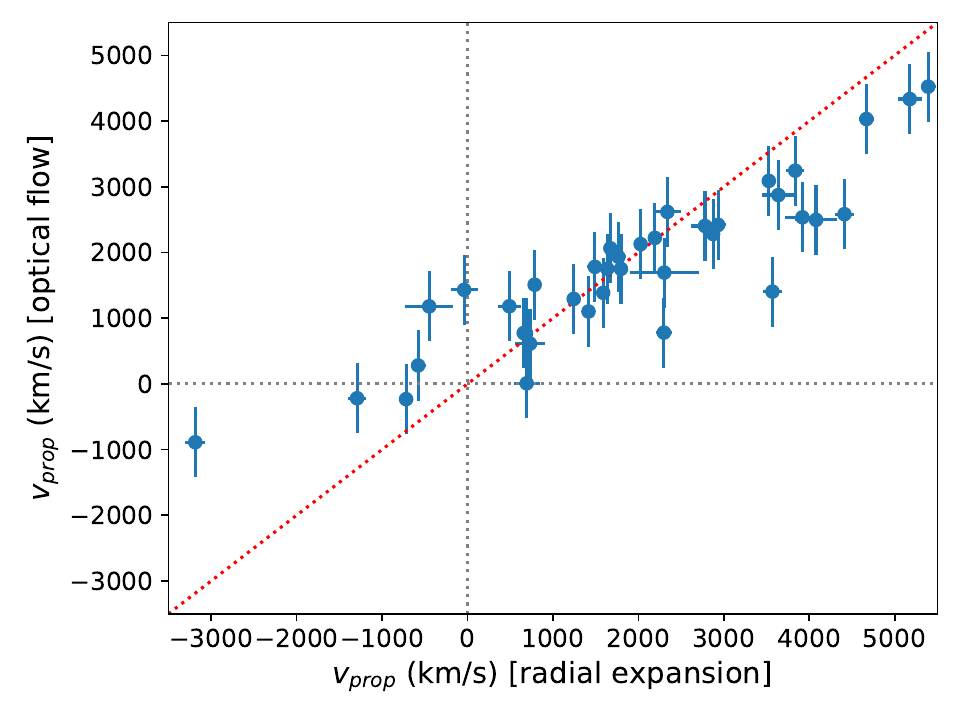}
    \caption{
   \label{fig:comparison} 
   Comparison of the proper motions based on the optical flow method (main text) with the radial expansion measurements (appendix). 
       {Alt text: Comparison of two proper motion measurements.}
    }
\end{figure}

Finally, we compare in Fig.~\ref{fig:comparison} 
the radial expansion results with the proper motions reported in the main text.
For the comparison we projected the optical flow measurements to the radial vector centered on the expansion center of Cas A as measured by \citet{thorstensen2001}.
Overall the comparison shows that the results are consistent. However, the radial expansion measurements find on average larger velocities, in particular
for $v_{\rm prop}\gtrsim 3000$~km\,s$^{-1}$, but also an outlier at $v_{\rm prop}\approx -3200$~km\,s$^{-1}$. For the northwestern pointing, toward the western part, 
for a few pixels both methods find inward motions, but these are larger in magnitude
for the radial velocity measurements. It has been previously reported that in these regions the reverse shock is moving backward \citep{sato2018,vink2022}.



\end{document}